\documentclass[12pt]{article}

\usepackage{amsfonts,amsmath,amssymb,accents,mathrsfs}
\usepackage[retainorgcmds]{IEEEtrantools}
\usepackage{multirow}
\usepackage{multicol}
\usepackage{tikz}
\usepackage{graphicx,color}
\usepackage{booktabs}
\usepackage{placeins}
\usepackage{XoohmE}
\usepackage[colorlinks,linkcolor=Blue,citecolor=Blue,bookmarks,bookmarksnumbered]{hyperref}
\usepackage{cite}
\usepackage[scaled=0.85]{helvet}
\usepackage[utf8]{inputenc}

\definecolor{Red}    {rgb}{0.90,0.00,0.12} %  1
\definecolor{Blue}   {rgb}{0.00,0.00,1.00} %  2
\definecolor{Green}  {rgb}{0.10,0.70,0.10} %  3
\definecolor{Turque} {rgb}{0.00,0.65,0.85} %  4
\definecolor{Orange} {rgb}{1.00,0.50,0.15} %  5
\definecolor{Magenta}{rgb}{1.00,0.00,1.00} %  6
\definecolor{Gold}   {rgb}{1.00,0.75,0.25} %  7
\definecolor{Seaweed}{rgb}{0.01,0.24,0.09} %  8
\definecolor{Purple} {rgb}{0.50,0.25,0.55} %  9
\definecolor{Brown}  {rgb}{0.43,0.26,0.32} % 10
\definecolor{grey1}  {rgb}{0.20,0.20,0.20} % 11
\definecolor{grey2}  {rgb}{0.40,0.40,0.40} % 12
\definecolor{grey3}  {rgb}{0.60,0.60,0.60} % 13
\definecolor{grey4}  {rgb}{0.80,0.80,0.80} % 14
\definecolor{grey5}  {rgb}{0.90,0.90,0.90} % 15

\def\a{{\alpha}}
\def\b{{\beta}}
\def\g{{\gamma}}
\def\d{{\delta}}
\def\r{{\rho}}

\def\t{{\tau}}

\def\th{{\theta}}
\def\S{{\Sigma}}
\def\L{{\Lambda}}

\def\ad{{\dot{\alpha}}}
\def\bd{{\dot{\beta}}}
\def\gd{{\dot{\gamma}}}

\def\thd{{\bar{\theta}}}

\def\M{{\mathcal{T}}}
\def\N{{\mathcal{J}}}
\def\S{{\mathcal{S}}}
\def\D{{\rm D}}
\def\Dd{{\bar{\rm D}}}
\def\pa{\partial}

\def\be{\begin{equation}}
\def\ee{\end{equation}}
\def\Ibea{\begin{IEEEeqnarray*}}
\def\Ieea{\end{IEEEeqnarray*}}
\def\n{\IEEEyesnumber}
\def\sn{\IEEEyessubnumber}

\newcommand{\bea}{\begin{eqnarray}}
\newcommand{\eea}{\end{eqnarray}}
\newcommand{\ena}{\end{eqnarray}}

\makeatletter
\def\section{\@startsection{section}{1}{\z@}
              {3ex plus-1ex minus-.2ex}{1pt plus1pt}{\large\sf\bfseries\boldmath}}
\def\subsection{\@startsection{subsection}{2}{\z@}
              {1.5ex plus-1ex minus-.2ex}{0.01pt plus1pt}{\sf\slshape}}
\def\subsubsection{\@startsection{subsubsection}{3}{\z@}
              {1.5ex plus-1ex minus-.2ex}{0.01pt plus0.2pt}{\sf\boldmath}}
\def\paragraph{\@startsection{paragraph}{4}{\z@}
              {.75ex \@plus.5ex \@minus.2ex}{-2mm}{\sf\bfseries\boldmath}}
\makeatother

\begin{document}
\thispagestyle{empty}
\noindent{\small
\hfill{HET-1722 {~} \\ % un-comment-out and specify when done}
$~~~~~~~~~~~~~~~~~~~~~~~~~~~~~~~~~~~~~~~~~~~~~~~~~~~~~~~~~~~~$
$~~~~~~~~~~~~~~~~~~~~\,~~~~~~~~~~~~~~~~~~~~~~~~~\,~~~~~~~~~~~~~~~~$
 {~}
}
\vspace*{8mm}
\begin{center}
{\large \bf
Higher Spin Superfield interactions with the Chiral Supermultiplet:
\\ [2pt]
Conserved Supercurrents and Cubic Vertices } \\   [12mm] {\large {
I. L. Buchbinder,\footnote{joseph@tspu.edu.ru}$^{a,b}$ S.\ James
Gates, Jr.,\footnote{sylvester${}_-$gates@brown.edu}$^{c}$ and
Konstantinos
Koutrolikos\footnote{kkoutrolikos@physics.muni.cz}$^{d}$ }}
\\*[10mm]
\emph{
\centering
$^a$Department of Theoretical Physics,Tomsk State Pedagogical University,\\
Tomsk 634041, Russia
\\[11pt]
$^b$National Research Tomsk State University,\\
Tomsk 634050, Russia
\\[11pt]
$^{c}$Department of Physics, Brown University,
\\[1pt]
Box 1843, 182 Hope Street, Barus \& Holley 545,
Providence, RI 02912, USA
\\[11pt]
and
\\[11pt]
$^d$ Institute for Theoretical Physics and Astrophysics, Masaryk University,
\\[1pt]
611 37 Brno, Czech Republic
}
 $$~~$$
  $$~~$$
 \\*[-8mm]
{ ABSTRACT}\\[4mm]
\parbox{142mm}{\parindent=2pc\indent\baselineskip=14pt plus1pt
We investigate cubic interactions between a chiral superfield and higher
spin superfield corresponding to irreducible representations of the 
$4D,\, \mathcal{N}=1$ super-Poincar\'{e}
algebra. We do this by demanding an invariance under the most general transformation,
linear in the chiral superfield. Following Noether's method
we construct an infinite tower of higher spin supercurrent multiplets
which are quadratic in the chiral superfield and include higher derivatives.  The
results are that a single, massless, chiral superfield can couple only to the half-integer
spin supermultiplets $(s+1,s+1/2)$ and for every value of spin there is an appropriate
improvement term that reduces the supercurrent multiplet to a minimal multiplet
which matches that of superconformal higher spins. On the other hand a single, massive, chiral
superfield can couple only to higher spin supermultiplets of type $(2l+2\hspace{0.3ex},\hspace{0.1ex}2l+3/2)$ and there is
no minimal multiplet. Furthermore, for the massless case we discuss
the component level higher spin currents and provide
explicit expressions for the integer and half-integer spin conserved currents together with a R-symmetry current.}
 \end{center}
$$~~$$
\vfill
\vfill
\clearpage
%
%%%%%%%%%%%%%%%%%%%%%%%%%%%%%%%%%%%%%%%%%%%%%%
%%%%%%%%% INTRODUCTION %%%%%%%%%%%%%%%%%%%%%%%
\section{Introduction}
Higher spin theories
\cite{hsh1,hsh2,hsh3,hsh4,hsh5,hsh6,hsh7,hsh8,hsh9} have a
considerable history and for a number of years drove the development
of many ideas in theoretical physics.  However, their role in
fundamental interactions is still not clear. On the one hand, all
the elementary particles observed in nature so far seem to
be concentrated in a region of spin values ($s$) such that $s\leq2$.
Moreover, this observation appears to be supported by a substantial
list of No-Go theorems
\cite{nogo1,nogo2,nogo3,nogo4,nogo5,nogo6,nogo7,nogo8,nogo9,nogo10,nogo11,nogo12,yesgo1,yesgo2}
(for reviews look in \cite{review1,review2}) suggesting that nature
stops with spin 2. On the other hand, if we want to understand
relativistic field theories and their quantum aspects in full
generality, there is no a priori reason to exclude higher spin
fields. In recent decades, this point was made undeniable due to the
crucial part that massless and massive higher spin particles play in
(\emph{i}) the softness of string interactions at high energy scales,
(\emph{ii}) the possibilities to describe string effects in the
framework of field theory, and (\emph{iii}) investigations of some
aspects of the holographic principle\footnote{For example,
Fradkin-Vasiliev cubic interaction vertex of massless higher spin
fields with gravity requires the \emph{AdS} background.}.

The construction of fully interacting higher spin theories is an
extremely exciting topic but also very difficult, mostly due to
the road blocks placed by the no-go results and maybe due to current
lack of (still
unknown) general principles. Also, one cannot exclude that higher
spin field theory is an effective theory for an underlying,
 so far unknown,
 more fundamental theory. Nevertheless, there are few
examples of successful approaches to higher spin theory such
as Vasiliev's theory \cite{hsh9,vt1,vt2,vt3} (for reviews look in
\cite{rvt1,rvt2,rvt3})
and the 3D Cherns-Simon higher spin-gravity formulation
\cite{cs1,cs2,cs3}. Despite their actual successes, these theories
still appear very complicated. For example, Vasiliev's theory provides an
infinite set of on-shell equations of motion and many conceptual
questions about observables, Lagrangian formulation, locality
\footnote{See e.g. \cite{Tar}, \cite{STar}.} and
quantization require continued study. In addition, the
Chern-Simons description of interacting higher spins is restricted to 3D
and has, in the massless case, no local degrees of freedom. Therefore, a lot of
the important questions concerning higher spin field theory are
still open\footnote{At present time, there is an extensive
literature on different aspects of higher spin field theory. For example see
the recent papers \cite{GKS,KMT,KT,BSZ,
M,DV,V,Z,BBB,BBF,BT1,BT2,SSZ,ST,T1,ST2} and references therein.}.

In higher spin theories the structure of possible interaction vertices
is essentially fixed by higher-spin symmetries. We
will consider the construction of the simplest vertices in the
supersymmetric higher-spin models. In this case, one can expect
that the supersymmetry will impose the additional restrictions on
the form of vertices and therefore one can hope to uncover
clarifications and simplifications in comparison to
non-supersymmetric higher-spin models.

The simplest higher spin interaction is described by the cubic
vertex. Therefore, we will begin with the construction of a
cubic vertex for supersymmetric field theory. It is well known that 
supersymmetric field models can be formulated on-shell in terms
of component fields or off-shell in terms of appropriate superfields
(see the text books \cite{GGRS}, \cite{BK}). Both these ways of
constructing supersymmetric field models have their own
advantages and disadvantages and complement each other. In this
paper we will follow the superfield approach which allows us to keep
manifest supersymmetry off-shell.

One kind of cubic interaction vertex for two types of fields can be
written in the form $jh,$ where $j$ is a current constructed from
fields of type $\phi$ (matter fields) and $h$ is a field of another
type (gauge fields). Because the gauge
field $h$ is defined up to gauge transformations, the current $j$
must satisfy some conservation laws, \emph{i.e.} it is conserved.
Higher-spin interactions on the base of conserved current have been
constructed and explored by many authors (see e.g.
\cite{current1,current2,current3,current4,current5,current6,current7,current8,current9,current10})
\footnote{A BRST approach to the construction of cubic vertex has been developed in \cite{BRSTcub-int}}.

In this work we will present the construction of the conserved
${\cal N}=1$ higher superspin supercurrent and supertrace that generate the cubic
interactions between super-Poincar\'{e} higher spin
supermultiplets which play the role of gauge fields and the chiral supermultiplet which will play the
role of matter. The higher spin supercurrent and higher
spin supertrace together constitute the higher spin supercurrent multiplet and are the corresponding analogues to the low-spin
supercurrent and supertrace of conventional supersymmetric theory (see \cite{GGRS}, \cite{BK}).

The strategy we follow is that of Noether's method, which is a
perturbative procedure that allows one to constrain the
allowed interactions by imposing invariance order by order in
the number of (super)fields. Such a treatment of interactions will
be very clear and useful for the cubic order. In our case the corresponding
transformation for the matter superfield is the most general 
transformation, consistent with its chiral nature and up to linear order 
terms in the superfield and for the higher spin superfields is their
gauge transformation. 

The paper is organized as follows. Section 2 is devoted to
discussing Noether's procedure and specific features of
$4D,~\mathcal{N}=1$ super-Poincar\'{e} higher spin theories. In section 3, we find the
most general transformation of chiral superfield up to linear order and observe
that the parameters of this transformation match the structure of the gauge transformations
of specific higher spin supermultiplets. Sections 4, 5 and 6 are
devoted to the construction of the higher spin supercurrent multiplet of a free massless
chiral and generate the cubic interactions with higher spins. We find that 
the massless chiral can be coupled only to higher spin supermultiplets
of type $(s+1,s+1/2)$. In section 7, we show that
for every value of integer $s$ there are two types of higher spin supercurrent multiplets,
the canonical and the minimal and one can go from the canonical to the minimal
by an appropriate choice of improvement terms. Furthermore, we demonstrate that the minimal multiplet
coincides with the supercurrent multiplet generated by superconformal higher spins. 
In section 8, we discuss the on-shell superspace conservation equations for
both supercurrent multiplets. For the case of minimal multiplet, we use the conservation equation
alone to derive a simpler expression for the higher spin supercurrent.
In section 9, we project to components and find explicit expressions for 
the spacetime conserved integer spin, half-integer spin and R-symmetry currents.
The integer spin current has two contributions, one of the boson - boson type that
matches the known expressions for the integer spin currents of a complex scalar and 
the other is of the fermion - fermion type which agrees with the known expressions of
integer spin currents of a spinor. The half-integer spin and R-symmetry currents, as far as we know,
appear in the literature for the first time. Section 10, is devoted to the massive chiral superfield.
We find that it can couple only to higher spin supermultiplets of type $(2l+2~,~2l+3)$
and we present new expressions for the higher spin supercurrent multiplet. For the massive chiral there
is no minimal multiplet. In Section 11, we summarize and discuss the results.

%%%%%%%%%%%%%%%%%%%%%%%%%%%%%%%%%%%%%
%%%%%%%%%%%%%% Noether %%%%%%%%%%%%%%%%%%
\section{Noether's method}\label{Nm}
In general, finding \emph{consistent} interactions is a very difficult problem
if there is no guiding principle. For the cases of spin 2 (GR) and spin 1 (YM)
there is a very well developed geometrical understanding (\emph{Riemannian
Manifolds} and \emph{Principle Bundles} respectively) that plays the role
of the guiding principle, but for higher spins we do not have this geometrical
input. In some extent, the geometrical interpretation of higher spin fields is
still mysterious. Therefore, we have to use alternative methods. The idea is
to relax any geometrical prejudice and have only algebraic requirements.  In
this case the physical guiding principle is that of gauge invariance and
\emph{consistent} interactions are the ones that are in agreement with
gauge symmetries. Keep in mind that this is a physical requirement in
order for the interacting theory to have the same degrees of freedom as
the free theory.

Noether's method is a systematic, perturbative, analysis of the
invariance requirement. In this approach one expands the action $S[
\phi,h]$ and the transformation of fields in a power series of a coupling
constant $g$
\Ibea{l}
S[\phi,h]=S_0[\phi]+gS_1[\phi,h]+g^2S_2[\phi,h]+\dots\n\\
\delta \phi=\delta_0[\xi]+g\delta_1[\phi,\xi]+g^2\delta_2[\phi,\xi]+\dots\n\\
\delta h=\delta_0[\zeta]+g\delta_1[h,\zeta]+g^2\delta_2[h,\zeta]+\dots\n
\Ieea
where $S_i[\phi,h]$ includes the interaction terms of order $i+2$ in the
number of fields and $\delta_i$ is the part of transformations with terms
of order $i$ in the number of fields.  Hence, invariance can now
be written iteratively up to the order we desire to investigate.  For the
free theory ($g^0$) and the cubic interactions ($g^1$), which is the first
step beyond free theory, invariance demands:
\Ibea{l} \n\label{cub-inv-constr}
g^0:~~\frac{\delta S_0}{\delta\phi}\delta_0\phi+\frac{\delta S_0}{\delta
h}\delta_0 h=0\sn\\
g^1:~~g\frac{\delta S_0}{\delta\phi}\delta_1\phi+g\frac{\delta S_1}{\delta
\phi}\delta_0\phi
+g\frac{\delta S_0}{\delta h}\delta_1 h+g\frac{\delta S_1}{\delta h}\delta_0
h=0\sn
\Ieea

In our case, the role of matter will be played by the chiral supermultiplet,
described by a chiral superfield $\Phi$ ($\Dd_{\ad}{\Phi}=0$). % and the
%free theory action is $S_{0}[\Phi,\bar{\Phi}]=\int d^8z \Phi\bar{\Phi}$.
At
the free theory level the chiral superfield does not have any gauge
transformation, $\delta_0\Phi=0$.

For the role of gauge fields we consider the massless, higher spin,
irreducible representations of the $4D,~\mathcal{N}=1$,
super-Poincar\'{e} algebra.
In the pioneer papers \cite{hsh5,Vas}, using a component formulation, 
free ${\cal N}=1$ supersymmetric massless higher spin models in
four dimensions have been constructed.  A superfield formulation 
was proposed in \cite{hss1,hss2,hss3} and further developed in subsequent papers \cite{hss4,hss5,hss6} and 
generalized by different authors\footnote{See also a formulation of
supersymmetric gauge theory in the framework of BRST approach
\cite{hssBRST}.}.
The results are\footnote{This is the ``economical'' description according to \cite{hss6}.} :
\begin{enumerate}
\item The integer superspin $Y=s$ supermultiplets $(s+1/2 , s)$ are described
by a pair of superfields
$\Psi_{\ad(s)\ad(s-1)}$ and $V_{\a(s-1)\ad(s-1)}$ with the following zero
order gauge transformations
\Ibea{l}\n\label{hstr1}
\d_0\Psi_{\ad(s)\ad(s-1)}=-\D^2L_{\a(s)\ad(s-1)}+\tfrac{1}{(s-1)!}\Dd_{(\ad_{
s-1}}\Lambda_{\a(s)\ad(s-2)}~,\sn\\  \d_0 V_{\a(s-1)\ad(s-1)}=\D^{\a_s}L_{\a(
s)\ad(s-1)}+\Dd^{\ad_s}\bar{L}_{\a(s-1)\ad(s)}\sn~.
\Ieea
%%%%
\item The half-integer superspin $Y=s+1/2$ supermultiplets $(s+1 , s+1/2)$
have two descriptions. One of them use the pair of superfields $H_{\a(s)
\ad(s)}$, $\chi_{\a(s)\ad(s-1)}$ with the following zero order gauge transformations
\Ibea{l}\n\label{hstr2}
\d_0 H_{\a(s)\ad(s)}=\tfrac{1}{s!}\D_{(\a_s}\bar{L}_{\a(s-1))\ad(s)}-\tfrac{1}{s!}
\Dd_{(\ad_s}L_{\a(s)\ad(s-1))}\sn\label{hstr2H}~,\\
\d_0\chi_{\a(s)\ad(s-1)}=\Dd^2L_{\a(s)\ad(s-1)}+\D^{\a_{s+1}}\Lambda_{\a(
s+1)\ad(s-1)}\sn
\Ieea
and the other one use the superfields $H_{\a(s)\ad(s)}$, $\chi_{\a(s-1)\ad(
s-2)}$ with
\Ibea{l}\n\label{hstr3}
\d_0 H_{\a(s)\ad(s)}=\tfrac{1}{s!}\D_{(\a_s}\bar{L}_{\a(s-1))\ad(s)}-\tfrac{1}{s!}
\Dd_{(\ad_s}L_{\a(s)\ad(s-1))}\sn~,\\
\d_0\chi_{\a(s-1)\ad(s-2)}=\Dd^{\ad_{s-1}}\D^{\a_s}L_{\a(s)\ad(s-1)}+\tfrac{
s-1}{s}\D^{\a_s}\Dd^{\ad_{s-1}}L_{\a(s)\ad(s-1)}
+\tfrac{1}{(s-2)!}\Dd_{(\ad_{s-2}}J_{\a(s-1)\ad(s-3))}~~\sn
\Ieea
\end{enumerate}
Consequently, the cubic interactions of the chiral superfield with the higher
spin multiplets, according to (\ref{cub-inv-constr}) must satisfy:
\Ibea{l}
\frac{\delta S_0}{\delta\Phi}\delta_1\Phi+\frac{\delta S_1}{\delta \mathcal{A
}}\delta_0 \mathcal{A}=0\n\label{Ncon}
\Ieea
where $\mathcal{A}$ is the set of superfields that participate in the description
of higher spin supermultiplets for any value of $s$.  In this language, the
collection of \emph{non-trivial} supercurrents that generate the cubic interaction
terms correspond to the terms $\frac{\delta S_1}{\delta \mathcal{A}}$. 
The word \emph{non-trivial} means
that (\emph{i}) the chiral superfield may not interact with all possible higher
spin supermultiplets (trivially zero supercurrents) and (\emph{ii}) for the ones
that it interacts with, we must check that these interactions can not be adsorbed
by redefinitions of the chiral superfield.

%%%%%%%%%%%%%%%%%%%%%%%%%%%%%%%%%%%%%%
%%%%%%%%%%%%%% delta \Phi %%%%%%%%%%%%%%%%%%%
\section{First order gauge transformation for chiral superfield}\label{dPhi}
In the previous section, we saw that the higher spin supercurrents of a
chiral superfield are controlled by $\delta_1\Phi$. That is the part of the
transformation of $\Phi$ which is linear in $\Phi$. Examples of transformations of this
type are generated by superdiffeomorphisms or the superconformal
group and have been used in the past \cite{Osborn,Magro} in order to
find the coupling of the chiral supermultiplet to supergravities.

In this section we present the higher spin version of this transformation.
The most general ansatz one can write for such a transformation
is\footnote{We use the conventions of \emph{Superspace}\cite{GGRS} which include $\left\{\D_{\a},\Dd_{\ad}\right\}=i\pa_{\a\ad}$, $\D^{\a}\D_{\a}=2\D^2$ and $\Dd^{\ad}\Dd_{\ad}=2\Dd^2$}:
\Ibea{ll}\n\label{tr}
\d_{g}\Phi=g\sum_{l=0}^{\infty}\sum_{k=0}^{\infty}&\left\{\vphantom{\frac12}
A^{\a(k+1)\ad(k)}_{l}~\Box^{l}~\D_{\a_{k+1}}\Dd_{\ad_k}\D_{\a_k}\dots\Dd_{\ad_1}\D_{\a_1}\Phi\right.\\
&+\Gamma^{\a(k)\ad(k+1)}_{l}~\Box^{l}~\Dd_{\ad_{k+1}}\D^2\Dd_{\ad_k}\D_{\a_k}\dots\Dd_{\ad_1}\D_{\a_1}\Phi\\
&+\Delta^{\a(k)\ad(k)}_{l}~\Box^{l}~\Dd_{\ad_k}\D_{\a_k}\dots\Dd_{\ad_1}\D_{\a_1}\Phi\\
&+\left.\vphantom{\frac12}E^{\a(k)\ad(k)}_{l}~\Box^{l}~\D^2\Dd_{\ad_k}\D_{\a_k}\dots\Dd_{\ad_1}\D_{\a_1}\Phi\right\}
\Ieea
and depends on four infinite families of coefficients $\{A_{\a(k+1)\ad(k)}^{l
},~\Gamma_{\a(k)\ad(k+1)}^{l},~\Delta_{\a(k)\ad(k)}^{l}, E_{\a(k)\ad(k)}^{
l}\}$ with independently symmetrized dotted and undotted indices.  To
make this transformation consistent with the chiral nature of $\Phi$ we
must have ($\Dd_{\bd}\d_{g}\Phi=0$):
\Ibea{l}\n\label{chiralconst}
A^{l}_{\a(k+1)\ad(k)}=-\tfrac{k+1}{k+2}~\Dd^{\ad_{k+1}}\Delta^{l}_{\a(k+1)\ad(k+1)}~,\sn
\vspace{1ex}\\
%%%%%%%%%%%%%%%%%%%%%%%%%%%%%%%%%%
\Gamma^{l}_{\a(k)\ad(k+1)}=\tfrac{1}{(k+1)!}~\Dd_{(\ad_{k+1}}\Delta^{l+1}_{\a(k)\ad(k))}~,\sn \vspace{1ex}\\
%%%%%%%%%%%%%%%%%%%%%%%%%%%%%%%%%%%
E^{l}_{\a(k)\ad(k)}=\Dd^2\Delta^{l+1}_{\a(k)\ad(k)}~,\sn \vspace{1ex}\\
%%%%%%%%%%%%%%%%%%%%%%%%%%%%%%%%%%%
\Dd_{(\bd}\Delta^0_{\a(k)\ad(k))}=0~,\sn\label{Impcon} \vspace{1ex}\\
\Dd_{\bd}\Delta^0=0~.\sn
\Ieea
The conclusion is that parameters $A_{\a(k+1)\ad(k)}^{l},~\Gamma_{\a(
k)\ad(k+1)}^{l},~E_{\a(k)\ad(k)}^{l}$ are not independent and furthermore
\Ibea{l}\n\label{uncp}
\Delta^0=\Dd^2\ell\sn\\
\Delta^0_{\a(k)\ad(k)}=\tfrac{1}{k!}\Dd_{(\ad_{k}}\ell_{\a(k)\ad(k-1))}~,\sn\\
\Delta^l_{\a(k)\ad(k)}~\text{is unconstrained for}~l\geq1\sn
\Ieea
where $\ell,~\ell_{\a(k)\ad(k-1)}$ are arbitrary.

From equation (\ref{Ncon}) it is evident that the parameters which
appear in the transformation of $\Phi$ must also appear in the zeroth
order gauge transformation of the higher spin superfields. Looking at
the gauge parameters that appear in (\ref{hstr1},\ref{hstr2},\ref{hstr3})
we find that there is no unconstrained parameter with the structure of
$\Delta^{l+1}_{\a(k)\ad(k)}$, but equations (\ref{hstr2}) and
(\ref{hstr3}) include unconstrained gauge parameters which match
the structure of $\ell_{\a(k)\ad(k-1)}$.
The conclusion is that in order to construct invariant theories
where the chiral superfield couples to purely higher spin supermultiplets
we have to consider the following transformation of $\Phi$:
\Ibea{ll}\n\label{tr2}
\d_{g}\Phi=&-g\sum_{k=0}^{\infty}\left\{\vphantom{\frac{1}{2}}~\Dd^2\ell^{\a(k+1)\ad(k)}~\D_{\a_{k+1}}\Dd_{\ad_k}\D_{\a_k}\dots\Dd_{\ad_1}\D_{\a_1}\Phi\right.\\
&\hspace{9ex}\left.\vphantom{\frac{1}{2}}-\tfrac{1}{(k+1)!}\Dd^{(\ad_{k+1}}\ell^{\a(k+1)\ad(k))}~\Dd_{\ad_{k+1}}\D_{\a_{k+1}}\dots\Dd_{\ad_1}\D_{\a_1}\Phi\right\}\\
&+g\Dd^2\ell~\Phi~\hspace{65ex}.
\Ieea
The last term of (\ref{tr2}) will generate coupling to the vector multiplet,
thus in order to consider purely higher spin interactions we should ignore it. However, for the sake of completeness we will not do that.

The second conclusion we can already reach, is that a theory of a
single chiral superfield can couple {\em {only}} to half-integer superspin
$Y=s+1/2$ supermultiplets. This is a consequence of the constraint
(\ref{Impcon}) whose solution matches the structure of the transformation of bosonic
superfields of half-integer superspin theories but crucially not that
of integer superspin.
%\footnote{This result may seem unexpected.
%However, in a future work a solution to this will be presented separately.}.

%%%%%%%%%%%%%%%%%%%%%%%%%%%%%%%%%%%%%%%%%%%%%%
%%%%%%%%%%%%%% Variation %%%%%%%%%%%%%%%%%%%%
\section[Variation of the action]{Constructing the Higher Spin Supercurrents I:\newline Varying the action}\label{VarAct}
Having found the appropriate first order transformation for the chiral superfield,
we use it to perform Noether's procedure for the cubic order terms, as described
in \textsection\ref{Nm} and construct the higher spin supercurrents of the chiral
supermultiplet.  We consider a free massless chiral superfield, so we start from the free action
\Ibea{l}
S_o=\int d^8z~\Phi\bar{\Phi}\n\label{S0}
\Ieea
and calculate its variation under $\delta_g\Phi$
\footnote{From this point forward, when the integration is over the entire superspace the measure $d^8z$ 
will not be explicitly written but it will be implied.}
:
\Ibea{l}\n
\d_gS_o=-g\int\sum_{k=0}^{\infty}\left\{\vphantom{\frac{1}{2}}\Dd^2\ell^{\a(k+1)\ad(k)}~\D_{\a_{k+1}}\Dd_{\ad_k}\D_{\a_k}\dots\Dd_{\ad_1}\D_{\a_1}\Phi~\bar{\Phi}~+c.c.\right.\\
\left.\hspace{18ex}-\tfrac{1}{(k+1)!}~\Dd^{(\ad_{k+1}}\ell^{\a(k+1)\ad(k))}~\Dd_{\ad_{k+1}}\D_{\a_{k+1}}\dots\Dd_{\ad_1}\D_{\a_1}\Phi~\bar{\Phi}~+c.c.
\vphantom{\frac{1}{2}}\right\}\\
\hspace{8ex}+g\int\left\{\Dd^2\ell+\D^2\bar{\ell}\right\}\Phi~\bar{\Phi}~\hspace{55ex}.
\Ieea
However, in the above expression we can freely add any pair of terms
$A_{\a(k+1)\ad(k)}$, $B_{\a(k+1)\ad(k+1)}$ such that they identically
satisfy the equation
\Ibea{l}\n
\Dd^2A_{\a(k+1)\ad(k)}=\Dd^{\ad_{k+1}}B_{\a(k+1)\ad(k+1)}~\hspace{5ex}.
\Ieea
These terms play the role of improvement terms. We can prove that there
are at least two pairs of them
\begin{enumerate}
\item $A_{\a(k+1)\ad(k)}=W_{\a(k+1)\ad(k)}$~,~
$B_{\a(k+1)\ad(k+1)}=\tfrac{k+1}{(k+2)(k+1)!}\Dd_{(\ad_{k+1}}W_{\a(
k+1)\ad(k))}~,$
\item $A_{\a(k+1)\ad(k)}=\tfrac{1}{(k+1)!}\D_{(\a_{k+1}}\Dd^{\ad_{k+1}}
\bar{U}_{\a(k))\ad(k+1)}$~,~
$B_{\a(k+1)\ad(k+1)}=\tfrac{1}{(k+1)!}\D_{(\a_{k+1}}\Dd^2\bar{U}_{\a(
k))\ad(k+1)}$
\end{enumerate}
which will be relevant for our discussion.
Hence, we can write for the variation of the $S_o$ action:
\Ibea{l}\n\label{dgS}
\delta_gS_o=-g\int\sum_{k=0}^{\infty}\left\{\vphantom{\frac{1}{2}} \Dd^2\ell^{\a(k+1)\ad(k)}~\M_{\a(k+1)\ad(k)}~+c.c.\right.\\
\left.\hspace{18ex}-\tfrac{1}{(k+1)!}~\Dd^{(\ad_{k+1}}\ell^{\a(k+1)\ad(k))}~\N_{\a(k+1)\ad(k+1)}~+c.c.\vphantom{\frac{1}{2}}\right\}\\
\hspace{8ex}+g\int\left\{\Dd^2\ell+\D^2\bar{\ell}\right\}~\N
\Ieea
where
\Ibea{l}\n
\M_{\a(k+1)\ad(k)}=\tfrac{1}{(k+1)!k!}\D_{(\a_{k+1}}\Dd_{(\ad_k}\D_{\a_k}\dots\Dd_{\ad_1)}\D_{\a_1)}\Phi~\bar{\Phi}+W_{\a(k+1)\ad(k)}\sn\label{M}\\
\hspace{14ex}+\tfrac{1}{(k+1)!}\D_{(\a_{k+1}}\Dd^{\ad_{k+1}}\bar{U}_{\a(k))\ad(k+1)}~,\\
{}\\
\mathcal{J}_{\a(k+1)\ad(k+1)}=\tfrac{1}{(k+1)!(k+1)!}\Dd_{(\ad_{k+1}}\D_{(\a_{k+1}}\dots\Dd_{\ad_1)}\D_{\a_1)}\Phi~\bar{\Phi}\sn\label{N}\\
\hspace{14ex}+\tfrac{1}{(k+1)!}\D_{(\a_{k+1}}\Dd^2\bar{U}_{\a(k))\ad(k+1)}+\tfrac{k+1}{(k+2)(k+1)!}\Dd_{(\ad_{k+1}}W_{\a(k+1)\ad(k))}~,\\
{}\\
\mathcal{J}=\Phi\bar{\Phi}\sn~\hspace{5ex}.
\Ieea

It is important to observe that these objects are not uniquely defined, but there is some freedom.
For example $\N$ is defined up to terms
$\D^{\a}\Dd^2\lambda_{\a}+\Dd^{\ad}\D^2\bar{\lambda}_{\ad}$ for an arbitrary $\lambda_{\a}$\footnote
{$\lambda_{\a}$ has its own redundancy $\lambda_{\a} \sim \lambda_{\a}+\Dd^{\ad}\zeta_{\a\ad}+i\D_{\a}\varrho$
with $\varrho=\bar{\varrho}$}, whereas $\mathcal{J}_{\a(k+1)\ad(k+1)}$ is defined up to terms $\Dd^{\ad_{k+2}}\Xi_{\a(k+1)\ad(k+2)}$.
Also
$\M_{\a(k+1)\ad(k)}$ has the freedom
\Ibea{ll}
\M_{\a(k+1)\ad(k)} \sim \M_{\a(k+1)\ad(k)} &+ \Dd_{(\ad_{k}}P^{(1)}_{\a(k+1)\ad(k-1))}+\Dd^{\ad_{k+1}}P^{(2)}_{\a(k+1)\ad(k+1)}\n\label{defFreedom}\\
&+\D_{(\a_{k+1}}\Dd^2R^{(1)}_{\a(k))\ad(k)}+\D^{\a_{k+2}}\Dd^2 R^{(2)}_{\a(k+2)\ad(k)}~.
\Ieea

Furthermore, equation (\ref{dgS}) points towards a coupling of the chiral with the first
formulation (\ref{hstr2}) of $(s+1,s+1/2)$ supermultiplets, but for that to
happen we must have $\mathcal{J}_{\a(k+1)\ad(k+1)}$ to be real. This is a
consequence of the reality of superfield $H_{\a(s)\ad(s)}$ and transformation
(\ref{hstr2H}). Thus, in order to couple the theory purely to half-integer
superspin multiplet, we must make sure that we can select the improvement
terms such that $\mathcal{J}_{\a(k+1)\ad(k+1)}
$=$\bar{\mathcal{J}}_{\a(k+1)\ad(k+1)}$. This will depend on the detailed structure of
the real and imaginary part of the term $\tfrac{1}{(k+1)!(k+1)!}\Dd_{(\ad_{
k+1}}\D_{(\a_{k+1}}\dots\Dd_{\ad_1)}\D_{\a_1)}\Phi~\bar{\Phi}$. The
investigation of these structures is the purpose of the following section.
Due to the chiral nature of $\Phi$, this term can be simply written as
$i^{k+1}\pa^{(k+1)}\Phi~\bar{\Phi}$, where for simpicity we omit the uncontracted indices and complete symmetrization of them
with appropriate symmetrization factors is understood. The symbol $\pa^{
(k)}$ denotes a string of $k$ spacetime derivatives.
%%%%%%%%%%%%%%%%%%%%%%%%%%%%%%%%%%%%%%%%%%%%%%
%%%%%%%%% S and T - ology %%%%%%%%%%%%%%%%%%%%
\section{The combinatorics of the imaginary part}
First of all, we decompose the quantity $i^{k+1}\pa^{(k+1)}\Phi~\bar{\Phi}$ to a real and an imaginary part
\Ibea{l}\n
i^{k+1}\pa^{(k+1)}\Phi~\bar{\Phi}=~\tfrac{i^{k+1}}{2}\left[\pa^{(k+1)}\Phi~\bar{\Phi}+(-1)^{k+1}\Phi~\pa^{(k+1)}\bar{\Phi}\right]\\
~~~~~~~~~~~~~~~~~~~~+\tfrac{i^{k+1}}{2}\left[\pa^{(k+1)}\Phi~\bar{\Phi}-(-1)^{k+1}\Phi~\pa^{(k+1)}\bar{\Phi}\right]
\Ieea
and then we focus at the imaginary part with the goal to clarify whether the various improvement terms
($W_{\a(k+1)\ad(k)}$, $U_{\a(k+1)\ad(k)}$) can modify it in order to
make $\mathcal{J}_{\a(k+1)\ad(k+1)}$ real. Notice the difference between even and odd values of $k+1$
\Ibea{l}\n
\mathcal{I}^{(k+1)}\equiv i\text{Im}[~i^{k+1}\pa^{(k+1)}\Phi~\bar{\Phi}~]=
\begin{cases}
\tfrac{i}{2}(-1)^l\left(\pa^{(2l+1)}\Phi~\bar{\Phi}+\Phi~\pa^{(2l+1)}\bar{\Phi}\right), ~\text{for}~k+1=2l+1,~l=0,1,\dots\\
{}\\
\tfrac{1}{2}(-1)^l\left(\pa^{(2l)}\Phi~\bar{\Phi}-\Phi~\pa^{(2l)}\bar{\Phi}\right), ~\text{for}~k+1=2l,~l=1,2,\dots
\end{cases}
\Ieea

The type of terms that appear above are a special case to the more general type $\pa^{(m)}\Phi~\pa^{(n)}\bar{\Phi}$ terms. It is easy to prove
that this type of terms satisfy the following recursion relations:
\Ibea{l}
\pa^{(m)}\Phi~\pa^{(n)}\bar{\Phi} = \pa\left( \pa^{(m-1)}\Phi~\pa^{(n)}\bar{\Phi} \right) - \pa^{(m-1)}\Phi~\pa^{(n+1)}\bar{\Phi}\n~~,\\
\pa^{(m)}\Phi~\pa^{(n)}\bar{\Phi} = \pa\left( \pa^{(m)}\Phi~\pa^{(n-1)}\bar{\Phi} \right) - \pa^{(m+1)}\Phi~\pa^{(n-1)}\bar{\Phi}\n~~.
\Ieea
Using these recursion formulas, one can prove that
\Ibea{l}
\pa^{(2l+1)}\Phi~\bar{\Phi}+\Phi~\pa^{(2l+1)}\bar{\Phi}=\sum_{n=0}^{l}c_{n}~\pa^{(2n+1)}\left\{ \pa^{(l-n)}\Phi~\pa^{(l-n)}\bar{\Phi} \right\}\n\\
\pa^{(2l)}\Phi~\bar{\Phi}-\Phi~\pa^{(2l)}\bar{\Phi}=\sum_{n=0}^{l-1}d_{n}~\pa^{(2n+1)}\left\{ \pa^{(l-n)}\Phi~\pa^{(l-n-1)}\bar{\Phi} - \pa^{(l-n-1)}\Phi~\pa^{(l-n)}\bar{\Phi} \right\}\n
\Ieea
with 
\Ibea{l}
c_{n}=(-1)^{l-n}\left[\binom{l+n+1}{l-n}+\binom{l+n}{l-n-1}\right]~,~d_n=(-1)^{l-n-1}\binom{l+n}{l-n-1}\n\label{coef}
\Ieea
These identities hold in general, not just for the chiral but for any (super)function $\Phi$. An alternative proof of them
can be found by expanding the right hand side using the identity
\Ibea{l}
\pa^{(m)}\left(A~B\right)=\sum_{i=0}^{m}\binom{m}{i}~\pa^{(m-i)}A~\pa^{i}B\n\label{derivdist}
\Ieea
and matching the coefficients of the various terms with those of the left hand side. Doing that, one will find the following consistency conditions
\Ibea{l}
\sum_{i=0}^{l}c_{i}\binom{2i+1}{l+i-p+1}=
\begin{cases}
1~\text{for}~ p=0\\
0~\text{for}~ p=1,2,\dots,l
\end{cases}
~,~\n\label{c}\\
\sum_{i=0}^{l-1}d_{i}\left[\binom{2i+1}{l+i-p+1}-\binom{2i+1}{l+i-p}\right]=
\begin{cases}
-1~\text{for}~ p=0\\
0~\text{for}~ p=1,2,\dots,l-1
\end{cases}\n\label{d}
\Ieea
which define the coefficients $c_n$, $d_n$ recursively and have (\ref{coef}) as solutions. Furthermore, due to (\ref{c}, \ref{d}) the coefficients
$c_{i}$ and $d_{i}$ also satisfy
\Ibea{l}
\sum_{i=0}^{l}c_{i}\binom{2i}{l-p+i}=(-1)^{p}~~,~~\sum_{i=0}^{l-1}d_{i}\left[\binom{2i}{l-p+i}-\binom{2i}{l-p+i-1}\right]=(-1)^{p+1}\n\label{cdcoeff}
\Ieea
\subsection[Odd Values]{Odd values of $k+1$}
With the above in mind, for the general odd case we get:
\Ibea{l}
\mathcal{I}^{2l+1}=\sum_{n=0}^{l}\tfrac{(-1)^l}{2}c_{n}~\pa^{(2n)}\left\{\D,\Dd\right\}\left[ \pa^{(l-n)}\Phi~\pa^{(l-n)}\bar{\Phi} \right]~,~ l=0,1,\dots\n
\Ieea
where using the supersymmetry algebra we have converted $i\pa$ to the anticommutator of the spinorial covariant derivatives. Notice that with
the exception of this part of the expression, everything else is real. So it will be beneficial if we convert the anticommutator of spinorial derivatives to a commutator of spinorial derivatives using the following identity,
\Ibea{l}
\left\{\D,\Dd\right\}=\left[\D,\Dd\right]+2\Dd\D\n\label{atoc}
\Ieea
The part with the commutator will be a real contribution and the left over term has the structure $\Dd\D\left(\dots\right)$. According to (\ref{N}) these terms can always be removed by an appropriate choice
of the improvement term $W_{\a(2l+1)\ad(2l)}$, thus the reality of $\N_{\a(2l+1)\ad(2l+1)}$ can always be guaranteed. Specifically we get:
\Ibea{l}
\mathcal{I}^{2l+1}=(-1)^{l}\sum_{n=0}^{l}c_{n}~\pa^{(2n)}\left[ \pa^{(l-n)}\D\Phi~\pa^{(l-n)}\Dd\bar{\Phi} \right]\n\\
~~~~~~~~~~-\tfrac{i}{2}(-1)^{l}\sum_{n=0}^{l}c_{n}~\pa^{(2n)}\left[ \pa^{(l-n+1)}\Phi~\pa^{(l-n)}\bar{\Phi} - \pa^{(l-n)}\Phi~\pa^{(l-n+1)}\bar{\Phi} \right]\\
~~~~~~~~~~+(-1)^{l}\sum_{n=0}^{l}c_{n}~\pa^{(2n)}\Dd\D\left[ \pa^{(l-n)}\Phi~\pa^{(l-n)}\bar{\Phi} \right]~~~.
\Ieea
The conclusion of this analysis is that the term $i^{2l+1}\pa^{(2l+1)}\Phi~\bar{\Phi}$ which appears in the expression of $\N_{\a(2l+1)\ad(2l+1)}$
can be written as:
\Ibea{l}\n
i^{2l+1}\pa^{(2l+1)}\Phi~\bar{\Phi}=X^{(2l+1)}_{\a(2l+1)\ad(2l+1)}+\tfrac{1}{[(2l+1)!]^2}\Dd_{(\ad_{2l+1}}\D_{(\a_{2l+1}}Z^{(2l+1)}_{\a(2l))\ad(2l))}
\Ieea
where
\Ibea{l}
X^{(2l+1)}_{\a(2l+1)\ad(2l+1)}= \tfrac{i}{2}(-1)^l\left[\pa^{(2l+1)}\Phi~\bar{\Phi}-\Phi~\pa^{(2l+1)}\bar{\Phi}\right]\n\\
~~~~~~~~~~~~~~~~~~~~~-\tfrac{i}{2}(-1)^{l}\sum_{n=0}^{l}c_{n}~\pa^{(2n)}\left[ \pa^{(l-n+1)}\Phi~\pa^{(l-n)}\bar{\Phi} - \pa^{(l-n)}\Phi~\pa^{(l-n+1)}\bar{\Phi} \right]\\
~~~~~~~~~~~~~~~~~~~~~+(-1)^{l}\sum_{n=0}^{l}c_{n}~\pa^{(2n)}\left[ \pa^{(l-n)}\D\Phi~\pa^{(l-n)}\Dd\bar{\Phi} \right]~~~,\\
Z^{(2l+1)}_{\a(2l)\ad(2l)}=(-1)^{l}\sum_{n=0}^{l}c_{n}~\pa^{(2n)}\left[ \pa^{(l-n)}\Phi~\pa^{(l-n)}\bar{\Phi} \right]\n
\Ieea
and both these quantities are real. These expressions can be further simplified using (\ref{derivdist}, \ref{cdcoeff}) to
\Ibea{l}
X^{(2l+1)}_{\a(2l+1)\ad(2l+1)}= 
i(-1)^{l}\sum_{p=1}^{2l} (-1)^{p}~\pa^{(p)}\Phi~\pa^{(2l+1-p)}\bar{\Phi}
+(-1)^{l}\sum_{p=0}^{2l} (-1)^{p}~\pa^{(p)}\D\Phi~\pa^{(2l-p)}\Dd\bar{\Phi}~,\n\label{Xodd}\\\\
Z^{(2l+1)}_{\a(2l)\ad(2l)}=(-1)^{l}\sum_{p=0}^{2l} (-1)^{p}~\pa^{(p)}\Phi~\pa^{(2l-p)}\bar{\Phi}\n\label{Zodd}
\Ieea
\subsection[Even Values]{Even values of $k+1$}
The same analysis can be done for the general even case. For that situation we get
\Ibea{l}
\mathcal{I}^{(2l)}=\tfrac{1}{2}(-1)^{(l-1)}\sum_{n=0}^{l-1}d_{n}~\pa^{(2n)}\left[ \pa^{(l-n+1)}\Phi~\pa^{(l-n-1)}\bar{\Phi}
-2\pa^{(l-n)}\Phi~\pa^{(l-n)}\bar{\Phi}+\pa^{(l-n-1)}\Phi~\pa^{(l-n+1)}\bar{\Phi} \right]\n\\
~~~~~~~~~+i(-1)^{(l-1)}\sum_{n=0}^{l-1}d_{n}~\pa^{(2n)}\left[ \pa^{(l-n)}\D\Phi~\pa^{(l-n-1)}\Dd\bar{\Phi}
-\pa^{(l-n-1)}\D\Phi~\pa^{(l-n)}\Dd\bar{\Phi} \right]\\
~~~~~~~~~+i(-1)^{(l-1)}\sum_{n=0}^{l-1}d_{n}~\pa^{(2n)}\Dd\D\left[ \pa^{(l-n)}\Phi~\pa^{(l-n-1)}\bar{\Phi}
-\pa^{(l-n-1)}\Phi~\pa^{(l-n)}\bar{\Phi} \right]~~~~.
\Ieea
Hence, the term $i^{2l}\pa^{(2l)}\Phi~\bar{\Phi}$ can be expressed in the following way:
\Ibea{l}\n
i^{2l}\pa^{(2l)}\Phi~\bar{\Phi}=X^{(2l)}_{\a(2l)\ad(2l)}+\tfrac{1}{[(2l)!]^2}\Dd_{(\ad_{2l}}\D_{(\a_{2l}}Z^{(2l)}_{\a(2l-1))\ad(2l-1))}
\Ieea
where
\Ibea{l}
X^{(2l)}_{\a(2l)\ad(2l)}=~\tfrac{1}{2}(-1)^{l}\left[\pa^{(2l)}\Phi~\bar{\Phi}+\Phi~\pa^{(2l)}\bar{\Phi}\right]\n\\
~~~~~~~~~~~~~~+\tfrac{1}{2}(-1)^{(l-1)}\sum_{n=0}^{l-1}d_{n}~\pa^{(2n)}\left[ \pa^{(l-n+1)}\Phi~\pa^{(l-n-1)}\bar{\Phi}
-2\pa^{(l-n)}\Phi~\pa^{(l-n)}\bar{\Phi}+\pa^{(l-n-1)}\Phi~\pa^{(l-n+1)}\bar{\Phi} \right]\\
~~~~~~~~~~~~~~+i(-1)^{(l-1)}\sum_{n=0}^{l-1}d_{n}~\pa^{(2n)}\left[ \pa^{(l-n)}\D\Phi~\pa^{(l-n-1)}\Dd\bar{\Phi}
-\pa^{(l-n-1)}\D\Phi~\pa^{(l-n)}\Dd\bar{\Phi} \right]~~,\\
Z^{(2l)}_{\a(2l-1)\ad(2l-1)}=i(-1)^{(l-1)}\sum_{n=0}^{l-1}d_{n}~\pa^{(2n)}\left[ \pa^{(l-n)}\Phi~\pa^{(l-n-1)}\bar{\Phi}
-\pa^{(l-n-1)}\Phi~\pa^{(l-n)}\bar{\Phi} \right]~~.
\Ieea
As in the previous case, both $X^{(2l)}_{\a(2l)\ad(2l)}$ and $Z^{(2l)}_{\a(2l-1)\ad(2l-1)}$ are real. Using (\ref{derivdist}, \ref{cdcoeff}) we can simplify these expressions further
\Ibea{l}
X^{(2l)}_{\a(2l)\ad(2l)}=
(-1)^{(l-1)}\sum_{p=1}^{2l-1} (-1)^{p}~\pa^{(p)}\Phi~\pa^{(2l-p)}\bar{\Phi}
+i(-1)^{l}\sum_{p=0}^{2l-1} (-1)^{p}~\pa^{(p)}\D\Phi~\pa^{(2l-1-p)}\Dd\bar{\Phi}~~,\n\label{Xeven}\\
Z^{(2l)}_{\a(2l-1)\ad(2l-1)}=i(-1)^{l}\sum_{p=0}^{2l-1} (-1)^{p}~\pa^{(p)}\Phi~\pa^{(2l-1-p)}\bar{\Phi}\n\label{Zeven}~~.
\Ieea
\section[Cubic Interactions and Supercurrents]{Constructing the Higher Spin Supercurrents II:\newline Gauge invariance and Cubic interactions}
The main point of the previous section is to prove that for every value of integer $m$
we can write
\Ibea{l}\n
i^{(k+1)}\pa^{(k+1)}\Phi~\bar{\Phi}=X^{(k+1)}_{\a(k+1)\ad(k+1)}+\tfrac{1}{[(k+1)!]^2}\Dd_{(\ad_{k+1}}\D_{(\a_{k+1}}Z^{(k+1)}_{\a(k))\ad(k))}\n\label{pureHS}
\Ieea
where $X^{(k+1)}_{\a(k+1)\ad(k+1)}$ and $Z^{(k+1)}_{\a(k)\ad(k)}$ are:
\Ibea{l}
X^{(k+1)}_{\a(k+1)\ad(k+1)}=
(-i)^{k-1}\sum_{p=1}^{k} (-1)^{p}~\pa^{(p)}\Phi~\pa^{(k+1-p)}\bar{\Phi}
+(-i)^{k}\sum_{p=0}^{k} (-1)^{p}~\pa^{(p)}\D\Phi~\pa^{(k-p)}\Dd\bar{\Phi}~~,~~\n\label{X}\\
Z^{(k+1)}_{\a(k)\ad(k)}=(-i)^{k}\sum_{p=0}^{k} (-1)^{p}~\pa^{(p)}\Phi~\pa^{(k-p)}\bar{\Phi}\n\label{Z}~~.
\Ieea
Thus the expression for $\N_{\a(k+1)\ad(k+1)}$ (\ref{N}) becomes:
\Ibea{ll}\n
\N_{\a(k+1)\ad(k+1)}&=X^{(k+1)}_{\a(k+1)\ad(k+1)}+\tfrac{1}{(k+1)!(k+1)!}\Dd_{(\ad_{k+1}}\D_{(\a_{k+1}}Z^{(k+1)}_{\a(k))\ad(k))}\label{N2}\\
&~+\tfrac{1}{(k+1)!}\D_{(\a_{k+1}}\Dd^2\bar{U}_{\a(k))\ad(k+1)}+\tfrac{k+1}{(k+2)(k+1)!}\Dd_{(\ad_{k+1}}W_{\a(k+1)\ad(k))}~~~.
\Ieea
This is useful because it makes obvious that we can always make $\N_{\a(k+1)\ad(
k+1)}$ real by choosing
\Ibea{ll}
W_{\a(k+1)\ad(k)}=&-\tfrac{k+2}{k+1}\D^2U_{\a(k+1)\ad(k)}-\tfrac{k+2}{k+1}\tfrac{1}{(k+1)!}\D_{(\a_{k+1}}Z^{(k+1)}_{\a(k))\ad(k)}\n~~.
\Ieea
With this choice we get
\Ibea{l}
\N_{\a(k+1)\ad(k+1)}=X^{(k+1)}_{\a(k+1)\ad(k+1)}
+\tfrac{1}{(k+1)!}\D_{(\a_{k+1}}\Dd^2\bar{U}_{\a(k))\ad(k+1)}-\tfrac{1}{(k+1)!}\Dd_{(\ad_{k+1}}\D^2U_{\a(k+1)\ad(k))}\n\label{N3}~~,\vspace{1ex}\\
\M_{\a(k+1)\ad(k)}=\tfrac{1}{(k+1)!}\D_{(\a_{k+1}}\M_{\a(k))\ad(k)}\n\label{M2}~~,
\vspace{1.2ex}\\
\M_{\a(k)\ad(k)}=i^k\pa^{(k)}\Phi~\bar{\Phi}-\tfrac{k+2}{k+1}Z^{(k+1)}_{\a(k)\ad(k)}+\tfrac{k+2}{k+1}\D^{\a_{k+1}}U_{\a(k+1)\ad(k)}+\Dd^{\ad_{k+1}}\bar{U}_{\a(k)\ad(k+1)}\n\label{M3}~~.
\Ieea
Due to equation (\ref{M2}), the variation of the action can be enhanced from
(\ref{dgS}) to the following, with the addition of the $\lambda_{\a(k+2)\ad(k)}$ term:
\Ibea{ll}\n\label{dsg2}
\d_gS_o=&-g\int\sum_{k=0}^{\infty}\left\{\vphantom{\frac{1}{2}}\left[\Dd^2\ell^{\a(k+1)\ad(k)}-\D_{\a_{k+2}}\lambda^{\a(k+2)\ad(k)}\right]\D_{\a_{k+1}}\mathcal{T}_{\a(k)\ad(k)}+c.c.\right.\\
&\hspace{12ex}\left.\vphantom{\frac{1}{2}}-\tfrac{1}{(k+1)!}\Dd^{(\ad_{k+1}}\ell^{\a(k+1)\ad(k))}~\mathcal{J}_{\a(k+1)\ad(k+1)}+c.c.\right\}\\
%%%%%%%%%%%%%%%%%%%%%%%%%%%%%%%%%%%%%%%
%%%%%%%%%%%%%%%%%%%%%%%%%%%%%%%%%%%%%%%
&+g\int\left\{\Dd^2\ell+\D^2\bar{\ell}\right\}\mathcal{J}~\hspace{49ex}.
\Ieea
In order to complete Noether's procedure and get an invariant theory
we have to add to the starting action $S_o$ the following higher spin, cubic
interaction terms
\Ibea{ll}\n\label{ci}
S_{\text{HS-$\Phi$ cubic interactions}}=&~g\int\sum_{k=0}^{\infty}\left\{\vphantom{\frac{1}{2}}~~
H^{\a(k+1)\ad(k+1)}~\mathcal{J}_{\a(k+1)\ad(k+1)}\right.\n\\
&\hspace{10ex}+\left.\chi^{\a(k+1)\ad(k)}~\D_{\a_{k+1}}\mathcal{T}_{\a(k)\ad(k)}+c.c.\vphantom{\frac{1}{2}}\right\}\\
%%%%%%%%%%%%%%%%%%%%%%%%%%%%%%%%%%%%%%%%%%%%%%%%%%%%%%%%%%%%%%%
%%%%%%%%%%%%%%%%%%%%%%%%%%%%%%%%%%%%%%%%%%%%%%%%%%%%%%%%%%%%%%%
&-g\int~V~\mathcal{J}
\Ieea
where $V$ is the real scalar superfield that describes the vector supermultiplet
and has the gauge transformation $\d_{0}V=\Dd^2\ell+\D^2\bar{\ell}$ and $H_{\a(k+1)\ad(k+1)},~\chi_{\a(k+1)\ad(k)}$ are the
superfields that describe the super-Poincar\'{e} higher spin $(k+2, k+3/2)$ supermultiplet with the gauge transformations of (\ref{hstr2}).
These cubic interaction terms generate the higher spin supercurrent $\N_{\a(k+1)\ad(k+1
)}$ and the higher spin supertrace $\M_{\a(k)\ad(k)}$.

As expected, the supercurrent $\N_{\a(k+1)\ad(k+1)}$ and supertrace $\M_{\a(k)\ad(k)}$
include higher derivative terms. This is a corollary of the \emph{Metsaev bounds}
\cite{Metsaev}, where the number of derivatives that appear in a non-trivial cubic
vertex is bounded from below by the highest spin involved and from above by
the sum of the spins involved. In our case, there is no upper bound on the spins
involved, which is consistent with the higher spin algebra structure\footnote{The
Jacobi identity requires an infinite tower of fields with unbounded spin}
\cite{Fradkin, Boulanger} thus making the number of derivatives that appear
in (\ref{ci}) unbounded (as in string field theory).

Due to the higher derivative terms and the fixed engineering dimensions of $H_{\a(k+1)\ad(k+1)},~\chi_{\a(k+1)\ad(k)}$
from the free theory of massless, super-Poincar\'{e} higher spins \cite{hss4,hss5,hss6}, we need to have an appropriate dimensionful
parameter $M$ in order to balance the engineering dimensions of (\ref{ci}
)\footnote{Multiply the terms inside the curly bracket with $\left(\tfrac{1}{M}\right)^{
k+1}$}, but since this effect can be easily tracked, for the sake of simplicity we
will not explicitly include it. However, it is important to remember its presence
since it introduces a scale into the theory. Also the parameter $M$ gives the connection
between the gauge parameters $\ell_{\a(k+1)\ad(k)}, \lambda_{\a(k+2)\ad(k)}$
that appear in (\ref{dsg2}) with the gauge parameters $L_{\a(k+1)\ad(k)},
\Lambda_{\a(k+2)\ad(k)}$ that appear in  (\ref{hstr2}).

The conclusion of this section is that a single chiral superfield
can have cubic interactions with only the half-integer superspin
supermultiplets $(s+1 , s+1/2)$ through the higher spin supercurrent and
supertrace that have been constructed above, but more importantly
although there are two possible descriptions of the $(s+1 , s+1/2)$
supermultiplet, the chiral superfield has a preference to only one
of them. The one that it chooses to interact with, is the one that
appears in the higher spin, $\mathcal{N}$=$2$ theories as presented
in \cite{Gates, hss4}.

%%%%%%%%%%%%%%%%%%%%%%%%%%%%%%%%%%%%%%
%%%%%%%%% Redefinitions %%%%%%%%%%%%%%%%%%%%%%
\section{Minimal multiplet of Noether higher spin supercurrents}\label{Redef}
In the previous section, we found explicit expressions for the higher spin
supercurrent and supertrace of the chiral superfield. Using the terminology of
\cite{Magro} these define the \emph{canonical} multiplet of Noether higher spin supercurrents
$\left\{\N_{\a(k+1)\ad(k+1)},\M_{\a(k)\ad(k)}\right\}$.
In this section we will show that for any value of the non-negative integer parameter $k$, there is
another higher spin supercurrent multiplet, called the \emph{minimal} multiplet
$\left\{\N^{\textit{min}}_{\a(k+1)\ad(k+1)},\M^{\textit{min}}_{\a(k)\ad(k)}\right\}$
and we arrive at it by an appropriate choice of the improvement terms such that
$\M^{\textit{min}}_{\a(k)\ad(k)}=0$.
In order to get some intuition about this process, it will be useful to examine first a simple example. 
\subsection{Coupling to Supergravity}\label{couplSUGRA}
For the case of $k=0$ the canonical multiplet of supercurrents we obtain is
\Ibea{l}\n
\N_{\a\ad}=\D_{\a}\Phi~\Dd_{\ad}\bar{\Phi}+\D_{\a}\Dd^2\bar{U}_{\ad}-\Dd_{\ad}\D^2U_{\a}~,\\
\M=-\Phi\bar{\Phi}+2\D^{\b}U_{\b}+\Dd^{\bd}\bar{U}_{\bd}
\Ieea
and they generate the cubic interactions between the chiral and
non-minimal supergravity supermultiplet. To investigate whether
$U_{\a}$ has the potential to completely eliminate one of these
supercurrents or reduce it to the point of being zero up to
redefinitions of $\Phi$, we consider the following
ansatz
\Ibea{l}\n
U_{\a}=f_1~\D_{\a}\Phi~\bar{\Lambda}+f_2~\Phi~\D_{\a}\bar{\Lambda}
\Ieea
where $\Lambda$ is the prepotential of the chiral field (\emph{i.e}
$\Phi$=$\Dd^2\Lambda$). It is straight forward to find that:
\Ibea{l}\n
\mathcal{J}_{\a\ad}=\left[1+2f_1-2f_2\right]~\D_{\a}\Phi~\Dd_{\ad}\bar{\Phi}-i[f_1-f_2]~\pa_{\a\ad}\Phi~\bar{\Phi}+i[f_1-f_2]~\Phi\pa_{\a\ad}\bar{\Phi}\n\\
~~~~~~~+[f_1-f_2]~\Dd_{\ad}\D_{\a}\left[\D^2\Phi~\bar{\Lambda}\right]-[f_1-f_2]~\D_{\a}\Dd_{\ad}\left[\Dd^2\bar{\Phi}~\Lambda\right]~,
\vspace{1ex}\\
\mathcal{T}=[-1+3f_2-3f_1]~\Phi\bar{\Phi}+2[f_1-f_2]~\D^2\Phi~\bar{\Lambda}+[f_1-f_2]~\Dd^2\bar{\Phi}~\Lambda\n\\
~~~~~~+2[f_1+f_2]~\D^2[\Phi~\bar{\Lambda}]+[f_1+f_2]\Dd^2[\bar{\Phi}~\Lambda]~~.
\Ieea
It is obvious that there is no choice of coefficients, $f_1$ and
$f_2$ that can make $\mathcal{T}$ vanish.  However, there
is a choice that makes $\M$ proportional to the zeroth order (free
theory) equation of motion of $\Phi$. This is important because
terms of this type can be absorbed by field redefinitions. If we
choose $-f_1=f_2=1/6$ we find
\Ibea{l}\n
\mathcal{J}_{\a\ad}=\tfrac{1}{3}\left\{\D_{\a}\Phi~\Dd_{\ad}\bar{\Phi}+i\pa_{\a\ad}\Phi~\bar{\Phi}-i\Phi\pa_{\a\ad}\bar{\Phi}\right\}+\tfrac{1}{3}\left[\D_{\a}\Dd_{\ad}(\Lambda\Dd^2\bar{\Phi})+c.c.\right]\\
\mathcal{T}=-\tfrac{2}{3}\D^2\Phi\bar{\Lambda}-\tfrac{1}{3}\Dd^2\bar{\Phi}\Lambda
\Ieea
and therefore by redefining $\Phi$ in the following manner
\Ibea{l}\n\label{phird}
\Phi\to\Phi+\tfrac{1}{3}g\Dd^2(\Lambda~\Dd^{\ad}\D^{\a}H_{\a\ad}) -\tfrac{1}{3}g\Dd^2(\Lambda~\D^{\a}\chi_{\a}) -\tfrac{2}{3}g\Dd^2(\Lambda~\Dd^{\ad}\bar{\chi}_{\ad})
\Ieea
the $S_o$ term will cancel the parts of the supercurrent and supertrace that
have a $\D^2\Phi,~\Dd^2\bar{\Phi}$ dependence. The outcome of this
procedure is the minimal multiplet of Noether supercurrent
for the case of supergravity
$\{\N^{\textit{min}}_{\a\ad},~\M^{\textit{min}} \}$,
which is
in agreement with the well known results in \cite{Ferrara,Magro}\footnote{Keep in mind the difference in conventions for the covariant spinorial derivatives.}
\Ibea{l}\n
\N^{\textit{min}}_{\a\ad}=\tfrac{1}{3}\left\{\D_{\a}\Phi\Dd_{\ad}\bar{\Phi}+i
\left( \pa_{\a\ad}\Phi \right) \, \bar{\Phi}-i\Phi \left(\pa_{\a\ad}\bar{\Phi}
\right) \, \right\}~,\sn\label{minimalSG}\\
\M^{\textit{min}}=0~.\sn
\Ieea
Furthermore, the cubic interaction of the chiral superfield with supergravity becomes
\Ibea{l}\label{sugraci}
S_{\text{SG-$\Phi$ cubic interactions}}=g\int
H^{\a\ad}~\N^{\textit{min}}_{\a\ad}\n~~~.
\Ieea
Nevertheless, we must keep in mind that $\Phi$'s redefinition
(\ref{phird}) will generate order $g^2$ terms which we ignore
because we focus on the cubic interaction terms. However, an interesting observation
is that part of these $g^2$ terms modify our
starting action $S_o$ in the following way
\Ibea{l}\n
\hspace{-3.8ex}\int \hspace{-0.6ex}\Phi\bar{\Phi}\to\hspace{-1ex}\int\hspace{-0.6ex}\left\{1-\tfrac{1}{9}g^2\left[\Dd^{\ad}\D^{\a}H_{\a\ad}-\D^{\a}\chi_{\a}-2\Dd^{\ad}\bar{\chi}_{\ad}\right]\left[\D^{\a}\Dd^{\ad}H_{\a\ad}+2\D^{\a}\chi_{\a}+\Dd^{\ad}\bar{\chi}_{\ad}\right]\right\}\hspace{-0.5ex}\Phi\bar{\Phi}
\Ieea
Of course this is nothing else than the perturbative construction
of the volume element as one should expect for a theory that
couples to supergravity.
%%%
%%%
\subsection{Coupling to Higher Superspin supermultiplets}
Based on the previous example, we should check whether the minimal multiplet exists for the general case or not.
According to (\ref{M3}), $\M_{\a(k)\ad(k)}$ is a linear combination of terms $\pa^{(p)}\Phi~\pa^{(k-p)}\bar{\Phi}$
for various values of the non-negative integer $p$. Therefore a relevant ansatz for the improvement term is:
\Ibea{l}
U^{(p)}_{\a(k+1)\ad(k)}=f^{(p)}_1\pa^{(p)}\D\Phi~\pa^{(k-p)}\bar{\Lambda}+f^{(p)}_2\pa^{(p)}\Phi~\pa^{(k-p)}\D\bar{\Lambda}\n
\Ieea
Following the instructions of (\ref{M3}) we calculate $\D^{\a_{k+1}}U^{(p)}_{\a(k+1)\ad(k)}$
\Ibea{ll}
\D^{\a_{k+1}}U^{(p)}_{\a(k+1)\ad(k)}&=f^{(p)}_2\tfrac{k+2}{k+1}~\pa^{(p)}\Phi~\pa^{(k-p)}\bar{\Phi}+f^{(p)}_1\tfrac{k+2}{k+1}~\pa^{(p)}\D^2\Phi~\pa^{(k-p)}\bar{\Lambda}\n\\
&~+f^{(p)}_2~\pa^{(p)}\D^{\a_{k+1}}\Phi~\pa^{(k-p)}\D\bar{\Lambda}-f^{(p)}_1~\pa^{(p)}\D\Phi~\pa^{(k-p)}\D^{\a_{k+1}}\bar{\Lambda}
\Ieea
To avoid potential confusion, the explicit expression of the term $\pa^{(p)}\D^{\a_{k+1}}\Phi~\pa^{(k-p)}\D\bar{\Lambda}$ is
\Ibea{l}
\tfrac{1}{(k+1)k!}\pa_{(\a_1(\ad_1}\dots\pa_{\a_p\ad_p}\D^{\a_{k+1}}\Phi~\pa_{\a_{p+1}\ad_{p+1}}\dots\pa_{\a_k\ad_k)}\D_{\a_{k+1})}\bar{\Lambda}
\Ieea
and by expanding the symmetrization of the indices, one can show that\newpage
\Ibea{l}
\pa^{(p)}\D^{\a_{k+1}}\Phi~\pa^{(k-p)}\D\bar{\Lambda}=\vspace{1ex}\\
\hspace{6ex}=-\tfrac{k-p+1}{k+1}~\pa^{(p)}\Phi~\pa^{(k-p)}\bar{\Phi}+i\tfrac{k-p}{k+1}~\pa^{(p)}\D\Phi~\pa^{(k-p-1)}\Dd\bar{\Phi}\n\vspace{1ex}\n\\
\hspace{7ex}~+i\tfrac{p}{k+1}~\pa^{(p-1)}\Dd\D^2\Phi~\pa^{(k-p)}\D\bar{\Lambda}
+i\tfrac{k-p}{k+1}~\pa^{(p)}\D^2\Phi~\pa^{(k-p-1)}\Dd\D\bar{\Lambda}
-\tfrac{1}{k+1}~\pa^{(p)}\D^2\Phi~\pa^{(k-p)}\bar{\Lambda}\vspace{1ex}\\
\hspace{7ex}~-i\tfrac{k-p}{k+1}\D^2\left[ \pa^{(p)}\Phi~\pa^{(k-p-1)}\Dd\D\bar{\Lambda} \right]
+\tfrac{1}{k+1}\D^2\left[ \pa^{(p)}\Phi~\pa^{(k-p)}\bar{\Lambda} \right]~~~~.
\Ieea
Similarly for the term $\pa^{(p)}\D\Phi~\pa^{(k-p)}\D^{\a_{k+1}}\bar{\Lambda}$ we get
\Ibea{l}
\pa^{(p)}\D\Phi~\pa^{(k-p)}\D^{\a_{k+1}}\bar{\Lambda}=\vspace{1ex}\\
\hspace{6ex}=~\tfrac{p+1}{k+1}~\pa^{(p)}\Phi~\pa^{(k-p)}\bar{\Phi}+i\tfrac{k-p}{k+1}~\pa^{(p)}\D\Phi~\pa^{(k-p-1)}\Dd\bar{\Phi}\n\vspace{1ex}\\
\hspace{7ex}+i\tfrac{p}{k+1}~\pa^{(p-1)}\Dd\D^2\Phi~\pa^{(k-p)}\D\bar{\Lambda}
-i\tfrac{k-p}{k+1}~\pa^{(p)}\D^2\Phi~\pa^{(k-p-1)}\D\Dd\bar{\Lambda}
+\tfrac{p+1}{k+1}~\pa^{(p)}\D^2\Phi~\pa^{(k-p)}\bar{\Lambda}\vspace{1ex}\\
\hspace{7ex}-i\tfrac{k-p}{k+1}\D^2\left[ \pa^{(p)}\D\Phi~\pa^{(k-p-1)}\Dd\bar{\Lambda} \right]
-\tfrac{p+1}{k+1}\D^2\left[ \pa^{(p)}\Phi~\pa^{(k-p)}\bar{\Lambda} \right]~~~~~.
\Ieea
Putting together all the above, we get
\Ibea{ll}
\D^{\a_{k+1}}U^{(p)}_{\a(k+1)\ad(k)}=&~
\tfrac{p+1}{k+1}\left(f^{(p)}_2-f^{(p)}_1\right)\pa^{(p)}\Phi~\pa^{(k-p)}\bar{\Phi}+i\tfrac{k-p}{k+1}\left(f^{(p)}_2-f^{(p)}_1\right)\pa^{(p)}\D\Phi~\pa^{(k-p-1)}\Dd\bar{\Phi}\n\\
&+\D^2\left[\vartheta\right]+\mathcal{O}(\D^2\Phi)
\Ieea
where $\D^2\left[\vartheta\right]$ is the sum of the terms that have the structure $\D^2\left[\dots\right]$ and $\mathcal{O}(\D^2\Phi)$ is the sum of the terms that depend on the combination $\D^2\Phi$.
Therefore the contribution of $U^{(p)}_{\a(k+1)\ad(k)}$ to $\M_{\a(k)\ad(k)}$ is
\Ibea{l}
\tfrac{k+2}{k+1}\D^{\a_{k+1}}U^{(p)}_{\a(k+1)\ad(k)}+\Dd^{\ad_{k+1}}\bar{U}^{(p)}_{\a(k)\ad(k+1)}=\vspace{1.5ex}\\
\hspace{0ex}=\hspace{-0.5ex}\tfrac{k+2}{k+1}\hspace{-0.6ex}\left(f^{(p)}_2-f^{(p)}_1\right)\hspace{-0.4ex}\pa^{(p)}\Phi~\pa^{(k-p)}\bar{\Phi}
+\tfrac{p+1}{k+1}\hspace{-0.4ex}\left(f^{(p)}_2-f^{(p)}_1\right)^{*}\hspace{-0.4ex}\pa^{(k-p)}\Phi~\pa^{(p)}\bar{\Phi}
-\tfrac{k-p}{k+1}\hspace{-0.4ex}\left(f^{(p)}_2-f^{(p)}_1\right)^{*}\hspace{-0.4ex}\pa^{(k-p-1)}\Phi~\pa^{(p+1)}\bar{\Phi}\\
\hspace{0ex}~+\tfrac{k+2}{k+1}\D^2\left[\vartheta\right]+\Dd^2\left[\bar{\vartheta}\right]+\tfrac{k+2}{k+1}\mathcal{O}(\D^2\Phi)+\bar{\mathcal{O}}(\Dd^2\bar{\Phi})+\D\zeta\n\label{ansatzU}
\Ieea
where we used $\pa^{(m)}\D\Phi~\pa^{(n)}\Dd\bar{\Phi}=\D\left(~\pa^{(m)}\Phi~\pa^{(n)}\Dd\bar{\Phi}~\right)-i\pa^{(m)}\Phi~\pa^{(n+1)}\bar{\Phi}$ and $\D\zeta$ is the sum of terms that have the structure $\D(\dots)$. It is important to observe that
due to (i) equation (\ref{M2}), (ii) the freedom in the definition of $\M_{\a(k+1)\ad(k)}$ (\ref{defFreedom})  and (iii) the freedom to redefine the chiral superfield in a manner similar to \textsection\ref{couplSUGRA}, all the terms in
the last line of (\ref{ansatzU}) can be ignored. Furthermore, the terms in the first line contribute to the appropriate terms of $\M_{\a(k)\ad(k)}$. Hence, if we consider
\Ibea{l}
U_{\a(k+1)\ad(k)}=\sum^{k}_{p=0}U^{(p)}_{\a(k+1)\ad(k)}\n\label{genU}
\Ieea
we have enough freedom to completely cancel $\M_{\a(k)\ad(k)}$. To illustrate this let us do this cancellation for $k=1$ and $k=2$ and then for the general case.
\begin{enumerate}
\item \underline{$k=1$}:  The canonical supertrace is $i\pa\Phi~\bar{\Phi}-\tfrac{3}{2}Z^{(2)}=-\tfrac{i}{2}~\pa\Phi~\bar{\Phi}+i\tfrac{3}{2}~\Phi~\pa\bar{\Phi}$~.\\
The contribution of $U^{(1)}$ is $\tfrac{3}{2}f^{(1)}~\pa\Phi~\bar{\Phi}+{f^{(1)}}^{*} \Phi~\pa\bar{\Phi}$, where $f^{(1)}=f^{(1)}_2-f^{(1)}_1$~.\\
The contribution of $U^{(0)}$ is $\tfrac{1}{2}{f^{(0)}}^*\pa\Phi~\bar{\Phi}+\left[\tfrac{3}{2}f^{(0)}-\tfrac{1}{2}{f^{(0)}}^*\right]\Phi~\pa\bar{\Phi}$, 
where $f^{(0)}=f^{(0)}_2-f^{(0)}_1$~.\\
We can cancel the supertrace competely if we select
\Ibea{l}
\tfrac{3}{2}f^{(1)}+\tfrac{1}{2}{f^{(0)}}^*=\tfrac{i}{2}\vspace{-2.1ex}\\
\hspace{35ex}\Rightarrow ~f^{(1)}=\tfrac{i}{10}~,~f^{(0)}=-\tfrac{7i}{10}\vspace{-2.1ex}\n\label{k=1}\\
{f^{(1)}}^*+\tfrac{3}{2}f^{(0)}-\tfrac{1}{2}{f^{(0)}}^*=-\tfrac{3i}{2}
\Ieea
\item \underline{$k=2$}: The canonical supertrace is $\tfrac{1}{3}\pa^{2}\Phi~\bar{\Phi}-\tfrac{4}{3}\pa\Phi~\pa\bar{\Phi}+\tfrac{4}{3}\Phi~\pa^{2}\bar{\Phi}$~.\\
The contribution of $U^{(2)}$ is $\tfrac{4}{3}f^{(2)}~\pa^{2}\Phi~\bar{\Phi}+{f^{(2)}}^*~\Phi~\pa^{2}\bar{\Phi}$, where $f^{(2)}=f^{(2)}_2-f^{(2)}_1$
~.\\
The contribution of $U^{(1)}$ is $\left[\tfrac{4}{3}f^{(1)} + \tfrac{2}{3}{f^{(1)}}^*\right]~\pa\Phi~\pa\bar{\Phi}-\tfrac{1}{3}{f^{(1)}}^*~\Phi~\pa^{2}\bar{\Phi}$, where $f^{(1)}=f^{(1)}_2-f^{(1)}_1$~.\\
The contribution of $U^{(0)}$ is $\tfrac{1}{3}{f^{(0)}}^*~\pa^{2}\Phi~\bar{\Phi}-\tfrac{2}{3}{f^{(0)}}^*~\pa\Phi~\pa\bar{\Phi}+\tfrac{4}{3}f^{(0)}~\Phi~\pa^{2}\bar{\Phi}$, where $f^{(0)}=f^{(0)}_2-f^{(0)}_1$~.\\
If we select
\Ibea{l}
\tfrac{4}{3}f^{(2)} +\tfrac{1}{3}{f^{(0)}}^*=-\tfrac{1}{3}\\
\tfrac{4}{3}f^{(1)} + \tfrac{2}{3}{f^{(1)}}^* -\tfrac{2}{3}{f^{(0)}}^*=\tfrac{4}{3}\hspace{2ex}\Rightarrow~ f^{(2)}=-\tfrac{1}{35}~,
~f^{(1)}=\tfrac{13}{35}~,~f^{(0)}=-\tfrac{31}{35}\n\label{k=2}\\
{f^{(2)}}^* -\tfrac{1}{3}{f^{(1)}}^* +\tfrac{4}{3}f^{(0)}=-\tfrac{4}{3}
\Ieea
then we completely cancel the supertrace.
\item \underline{General $k$}: For the general case, using (\ref{genU}) we can show that up to terms that can be ignored due to chiral redefinition and the freedom in the definitions of the supertrace (\ref{defFreedom}, \ref{M2}) we get:
\Ibea{l}
\tfrac{k+2}{k+1}\D^{\a_{k+1}}U_{\a(k+1)\ad(k)}+\Dd^{\ad_{k+1}}\bar{U}_{\a(k)\ad(k+1)}=\vspace{1.5ex}\\
=\left\{\tfrac{k+2}{k+1}f^{(k)}+\tfrac{1}{k+1}{f^{(0)}}^{*}\right\}\pa^{(k)}\Phi~\bar{\Phi}+
\sum_{p=0}^{k-1}\left\{\tfrac{k+2}{k+1}f^{(p)}+\tfrac{k+1-p}{k+1}{f^{(k-p)}}^{*}-\tfrac{p+1}{k+1}{f^{(k-1-p)}}^{*}\right\}~\pa^{(p)}\Phi~
\pa^{(k-p)}\bar{\Phi}
\Ieea
where $f^{(p)}=f^{(p)}_2-f^{(p)}_1$.
Then in order to cancel the supertrace, according to (\ref{Z}, \ref{M3}) we must have
\Ibea{l}\n
(k+2)f^{(k)}+{f^{(0)}}^{*}=(i)^{k}\sn~,~~\\
(k+2)f^{(p)}+(k+1-p){f^{(k-p)}}^*-(p+1){f^{(k-1-p)}}^*=(-1)^{k+p}~(i)^{k}~(k+2)~,\sn\\
\hspace{58ex}p=0,1,\dots,k-1~.
\Ieea
This is a system of $k+1$ linear equations for the $k+1$ parameters $f^{(p)},~p=0,1,...,k$.
The solution is
\Ibea{l}
f^{(p)}=(-1)^{k+p}~(i)^{k}~\frac{\sum\limits_{j=0}^{k-p}\binom{k+j+1}{p+j+1}\binom{k+1-j}{p+1}}{\binom{2k+3}{k+2}}~,~~p=0,1,...,k\n
\Ieea
\end{enumerate}
The result is that for any value of $k$, we can find an improvement term in order to go to the minimal multiplet of higher spin supercurrents $\{~\N^{\textit{min}}_{\a(k+1)\ad(k+1)}~,~\M^{\textit{min}}_{\a(k)\ad(k)}\}$ where
\Ibea{l}
\N^{\textit{min}}_{\a(k+1)\ad(k+1)}=if^{(k)}~\pa^{(k+1)}\Phi~\bar{\Phi}-i{f^{(k)}}^{*}~\Phi~\pa^{(k+1)}\bar{\Phi}\n\label{minimalcurrent}\\
\hspace{16ex}
+i\sum_{p=1}^{k}\left\{ (-1)^{k+p}~(i)^{k}+f^{(p-1)}-{f^{(k-p)}}^{*}\right\} ~\pa^{(p)}\Phi~\pa^{(k+1-p)}\bar{\Phi}\\
\hspace{16ex}
+\sum_{p=0}^{k}\left\{ (-1)^{k+p}~(i)^{k}-f^{(p)}-{f^{(k-p)}}^{*}\right\}~\pa^{(p)}\D\Phi~\pa^{(k-p)}\Dd\bar{\Phi}~~,\\
\M^{\textit{min}}_{\a(k)\ad(k)}=0\n~~~~.
\Ieea
For $k=1$ and $k=2$ we get
\Ibea{l}
\N^{\textit{min}}_{\a\b\ad\bd}=-\tfrac{1}{10}~\pa^{(2)}\Phi~\bar{\Phi}-\tfrac{1}{10}~\Phi~\pa^{(2)}\bar{\Phi}+\tfrac{2}{5}~\pa\Phi~\pa\bar{\Phi}
-\tfrac{1}{5}i~\D\Phi~\pa\Dd\bar{\Phi}+\tfrac{1}{5}i~\pa\D\Phi~\Dd\bar{\Phi}\n\label{minimalk=1}\vspace{2ex}~~,\\
\N^{\textit{min}}_{\a\b\g\ad\bd\gd}=-\tfrac{i}{35}~\pa^{(3)}\Phi~\bar{\Phi}+\tfrac{i}{35}~\Phi~\pa^{(3)}\bar{\Phi}
+i\tfrac{9}{35}~\pa^{(2)}\Phi~\pa\bar{\Phi}-i\tfrac{9}{35}~\pa\Phi~\pa^{(2)}\bar{\Phi}\n\label{minimalk=2}\\
~~~~~~~~~~~~~~-\tfrac{3}{35}~\pa^{(2)}\D\Phi~\Dd\bar{\Phi}-\tfrac{3}{35}~\D\Phi~\pa^{(2)}\Dd\bar{\Phi}+\tfrac{9}{35}~\pa\D\Phi~\pa\Dd\bar{\Phi}~~.
\Ieea
These expressions match the results of \cite{KMT} which give the superconformal higher spin supercurrent.
In the minimal supercurrent multiplet, the cubic interactions of the chiral supermultiplet with the higher spin supermultiplets are
\Ibea{ll}\n\label{minci}
S_{\text{HS-$\Phi$ cubic interactions}}=&~g\int\sum_{k=0}^{\infty}~~
H^{\a(k+1)\ad(k+1)}~\N^{\textit{min}}_{\a(k+1)\ad(k+1)}\n~~~~.
\Ieea
%%%%%%%%%%%%%%%%%%%%%%%%%%%%%%%%%%
%%%%%%%%% Conservation Equation %%%%%%%%%%%%%
\section{On-shell Conservation equations}\label{secConEq}
Using Noether's method, we have constructed an invariant action
up to order $g$. Hence, for every unconstrained parameter $\ell_{\a(k+1)\ad(k)}$ and $\ell$
we generate a Bianchi identity, which express the invariance of the action.
Once we go on-shell and take into account the equation of motion
of $\Phi$,
the Bianchi identities reduce
to the following on-shell conservation equations for the \emph{canonical} multiplet of the higher spin supercurrents. 
\Ibea{l}
\Dd^{\ad_{k+1}}\N_{\a(k+1)\ad(k+1)}=\tfrac{1}{(k+1)!}
\Dd^2\D_{(\a_{k+1}}\M_{\a(k))\ad(k)}~,~k=0,1,2,\dots\n
\label{ce1}\\
%%%%%%%%%%%%%%%%%%%%%
\Dd^2\N=0\sn\label{ce2}\n
\Ieea
It is straightforward to verify the validity of these on-shell equations
using the expressions (\ref{N3}, \ref{M2}, \ref{M3}).

For the \emph{minimal} multiplet, the conservation equation takes the much simpler form
\Ibea{l}
\Dd^{\ad_{k+1}}\N^{\textit{min}}_{\a(k+1)\ad(k+1)}=0~,~k=0,1,2,\dots\n
\label{mince}
\Ieea
After a bit of work, one can verify that equation (\ref{minimalcurrent}) satisfies this conservation equation. However, instead of using (\ref{minimalcurrent}) we can get a simpler expression for the minimal higher spin supercurrent by using the conservation equation to define the coefficients of the various terms. From the previous section we know that the general ansatz for the minimal, higher spin supercurrent is
\Ibea{l}
\N^{\textit{min}}_{\a(s)\ad(s)}=\sum_{p=0}^{s}~a_{p}~\pa^{(p)}\Phi~\pa^{(s-p)}\bar{\Phi}+
\sum_{p=0}^{s-1}~b_{p}~\pa^{(p)}\D\Phi~\pa^{(s-p-1)}\Dd\bar{\Phi}\n~~~~.
\Ieea
We also know that $\N^{\textit{min}}_{\a(s)\ad(s)}$ must be real, hence
\Ibea{l}
a_{p}=a^{*}_{s-p}~~,~~p=0,1,...,s\n\label{sc1}~~,\\
b_{p}=b^{*}_{s-p-1}~~,~~p=0,1,...,s-1\label{sc2}\n
\Ieea
and the on-shell conservation (~$\Dd^{\ad_{s}}\N^{\textit{min}}_{\a(s)\ad(s)}=0$~), also gives the constraint
\Ibea{l}
i~a_{p+1}\left[\vphantom{\frac12}\tfrac{p+1}{s}\right]+b_{p}\left[\vphantom{\frac12}\tfrac{s-p}{s}\right]=0~~,~~p=0,1,...,s-1\n\label{sc3}~~.
\Ieea
The constraints (\ref{sc1}, \ref{sc2}, \ref{sc3}) fix $a_{p}$ and $b_{p}$ to be (up to a real proportionality constant)
\Ibea{l}
a_{p}=(-1)^{p}(i)^{s}~\binom{s}{p}^2\n~~,\\
b_{p}=(-1)^{p}(i)^{s+1}~\left(\frac{s-p}{p+1}\right)\binom{s}{p}^2\n
\Ieea
and $\N^{\textit{min}}_{\a(s)\ad(s)}$ is proportional to
\Ibea{l}
\N^{\textit{min}}_{\a(s)\ad(s)}\sim(i)^{s}\sum_{p=0}^{s}~(-1)^{p}\binom{s}{p}^2\left\{~\pa^{(p)}\Phi~\pa^{(s-p)}\bar{\Phi}+
i\left(\frac{s-p}{p+1}\right)~\pa^{(p)}\D\Phi~\pa^{(s-p-1)}\Dd\bar{\Phi}~\right\}\n\label{FCtemp}~~~~.
\Ieea
We can fix the overall constant of proportionality by comparing this expression to (\ref{minimalcurrent}), thus we get
\Ibea{l}
\N^{\textit{min}}_{\a(s)\ad(s)}=\frac{(-i)^{s}}{\binom{2s+1}{s+1}}\sum_{p=0}^{s}~(-1)^{p}\binom{s}{p}^2\left\{~\pa^{(p)}\Phi~\pa^{(s-p)}\bar{\Phi}+
i\left(\frac{s-p}{p+1}\right)~\pa^{(p)}\D\Phi~\pa^{(s-p-1)}\Dd\bar{\Phi}~\right\}\n\label{FC}~~~~.
\Ieea
It is easy to check that this expression agrees with equations (\ref{minimalSG}, \ref{minimalk=1}, \ref{minimalk=2}, \ref{minimalcurrent}) and up to an overall coefficient it also agrees with the results in \cite{KMT}.
%%%%%%%%%%%%%%%%%%%%%%%%%%%%%%%%%%%%%%%%
%%%%%%%%% Component Discussion %%%%%%%%%%%%%%%%%%%
\section{Component discussion}\label{components}
In the literature there are various sets of conserved currents that
generate the cubic interactions of a complex scalar (two spin $0$)
and a spinor (one spin $1/2$) with higher spins
\cite{current1,current2,current3,current4,current5,current6}. It is
important to find how the results of previous sections translate
at the component description.

In principle, we can start with equation (\ref{minci}) and switch to the
component formulation by evaluating the $\theta$ integrals in order to
find the component analogue.  However, for the purpose of identifying
the higher spin, conserved currents, a conceptual cleaner approach
would be to start with the superspace conservation equation (\ref{mince}) and project it
down to the component level, in order to derive the spacetime
conservation equation of the currents. The latter is the approach
that we will follow and the definition of components we will use is:
\Ibea{ll}\n
\Phi^{(0,0)}_{\alpha(n)\dot{\alpha}(m)}=\Phi_{\alpha(n)\dot{\alpha}(m)}|_{\th=0}~~
,&\hspace{-9ex}\Phi^{(1,0)}_{\beta\alpha(n)\dot{\alpha}(m)}=D_{\beta}\Phi_{\alpha(n)\dot{\alpha}(m)}|_{\th=0}~,\\
\Phi^{(0,1)}_{\alpha(n)\dot{\beta}\dot{\alpha}(m)}=\bar{D}_{\dot{\beta}}\Phi_{\alpha(n)\dot{\alpha}(m)}|_{\th=0}~~
,&\hspace{-9ex}\Phi^{(1,1)}_{\beta\alpha(n)\dot{\beta}\dot{\alpha}(m)}=-\tfrac{1}{2}\left[D_{\beta},\bar{D}_{\dot{\beta}}\right]\Phi_{\alpha(n)\dot{\alpha}(m)}|_{\th=0}~,\\
\Phi^{(2,0)}_{\alpha(n)\dot{\alpha}(m)}=-D^2\Phi_{\alpha(n)\dot{\alpha}(m)}|_{\th=0}~~
,&\hspace{-9ex}\Phi^{(0,2)}_{\alpha(n)\dot{\alpha}(m)}=-\bar{D}^2\Phi_{\alpha(n)\dot{\alpha}(m)}|_{\th=0}~,\\
\Phi^{(2,1)}_{\alpha(n)\dot{\beta}\dot{\alpha}(m)}=-\tfrac{1}{2}\left\{D^2,\bar{D}_{\dot{\beta}}\right\}\Phi_{\alpha(n)\dot{\alpha}(m)}|_{\th=0}~~
,&\hspace{-9ex}\Phi^{(1,2)}_{\beta\alpha(n)\dot{\alpha}(m)}=-\tfrac{1}{2}\left\{\bar{D}^2,D_{\beta}\right\}\Phi_{\alpha(n)\dot{\alpha}(m)}|_{\th=0}~,\\
\Phi^{(2,2)}_{\alpha(n)\dot{\alpha}(m)}=\tfrac{1}{2}\{D^2,\bar{D}^2\}\Phi_{\alpha(n)\dot{\alpha}(m)}|-\tfrac{1}{4}\Box\Phi_{\alpha(n)\dot{\alpha}(m)}|_{\th=0}~~~~.
\Ieea
The various components are labeled by the name of the superfield
they come from and their position $(n,m)$ in its $\theta$ expansion
\Ibea{ll}
\Phi_{\a(n)\ad(m)}=&~\Phi_{\a(n)\ad(m)}+\th^{\b}\Phi^{(1,0)}_{\b\a(n)\ad(m)}+\thd^{\bd}\Phi^{(0,1)}_{\a(n)\bd\ad(m)}+\th^{2}\Phi^{(2,0)}_{\a(n)\ad(m)}
+\thd^{2}\Phi^{(0,2)}_{\a(n)\ad(m)}\n\\
&\hspace{-1ex}+\th^{\b}\thd^{\bd}\Phi^{(1,1)}_{\b\a(n)\bd\ad(m)}+\th^{\b}\thd^{2}\Phi^{(1,2)}_{\b\a(n)\ad(m)}+\th^{2}\thd^{\bd}\Phi^{(2,1)}_{\a(n)\bd\ad(m)}
+\th^{2}\thd^{2}\Phi^{(2,2)}_{\a(n)\ad(m)}~~.
\Ieea
Furthermore, components with more than one index of the same type
can be decomposed into symmetric (S) and anti-symmetric (A) pieces
as follows
\Ibea{l}\n
F_{\b\a(n)\ad(m)}=F^{(S)}_{\b\a(n)\ad(m)}+\tfrac{n}{(n+1)!}C_{\b(\a_n}F^{(A)}_{\a(n-1))\ad(m)}~~,\\
F^{(S)}_{\b\a(n)\ad(m)}=\tfrac{1}{(n+1)!}F_{(\b\a(n))\ad(m)}~,~
F^{(A)}_{\a(n-1)\ad(m)}=C^{\b\a_n}F_{\b\a(n)\ad(m)}~~.
\Ieea

Using the above, it is straightforward to project equation (\ref{mince}) and the results we find 
for the bosonic components are:
\Ibea{l}\n
\pa^{\a_{s}\ad_{s}}\N^{\textit{min}~(0,0)}_{\a(s)\ad(s)}=0\sn\label{00}~~,\vspace{1.4ex}\\
\N^{\textit{min}~(0,2)}_{\a(s)\ad(s)}=0~~,\sn\vspace{1.4ex}\\
\N^{\textit{min}~(1,1)(S,A)}_{\a(s+1)\ad(s-1)}=-\tfrac{i}{2(s+1)!}\pa_{(\a_{s+1}}{}^{\ad_{s}}\N^{\textit{min}~(0,0)}_{\a(s))\ad(s)}\sn\label{11SA}~~,
\vspace{1.4ex}\\
\N^{\textit{min}~(1,1)(A,A)}_{\a(s-1)\ad(s-1)}=0\sn~~,\vspace{1.4ex}\\
\pa^{\a_{s+1}\ad_{s+1}}\N^{\textit{min}~(1,1)(S,S)}_{\a(s+1)\ad(s+1)}=0\sn\label{b}~~,\vspace{1.4ex}\\
\N^{\textit{min}~(2,2)}_{\a(k+1)\ad(k+1)}=-\tfrac{1}{4}\Box\N^{(0,0)}_{\a(k+1)\ad(k+1)}\sn
\Ieea
and for the fermionic components we get:
\Ibea{l}\n
\N^{\textit{min}~(0,1)(A)}_{\a(s)\ad(s-1)}=~0~~,\sn\label{01A}\vspace{1.4ex}\\
%
%%%%%%%%%%%%%%%%%%%%%%%%%%%%%%%%%%%%%%%%%%%%%%%%%%%%%%%%%%%%%%%%%%%%%%%%%%%%%%%%%%%%%%%%%%%%%%%
\N^{\textit{min}~(1,2)(S)}_{\a(s+1)\ad(s)}=~~\tfrac{i}{2(s+1)!}\pa_{(\a_{s+1}}{}^{\ad_{s+1}}\N^{\textit{min}~(0,1)(S)}_{\a(s))\ad(s+1)}\sn~~,\vspace{1.4ex}\\
%
%%%%%%%%%%%%%%%%%%%%%%%%%%%%%%%%%%%%%%%%%%%%%%%%%%%%%%%%%%%%%%%%%%%%%%%%%%%%%%%%%%%%%%%%%%%%%%%
\N^{\textit{min}~(1,2)(A)}_{\a(s-1)\ad(s)}=~0\sn~~,\vspace{1.4ex}\\
\pa^{\a_{s+1}\ad_{s}}\N^{\textit{min}~(1,0)(S)}_{\a(s+1)\ad(s)}=~0~~.\sn\label{f}
\Ieea
The lesson is that the component $\N^{\textit{min}~(1,1)(S,S)}_{\a(s+1)\ad(s+1)}$ is the \emph{minimal} integer spin current and equation (\ref{b}) is its conservation equation. The cubic interactions it generates are of the type
\Ibea{l}\n
\int d^4x \sum_{s=0}^{\infty} h^{\a(s+1)\ad(s+1)}\N^{\textit{min}~(1,1)(S,S)}_{\a(s+1)\ad(s+1)}
\Ieea
where the field $h_{\a(s+1)\ad(s+1)}$ is the symmetric, traceless part of
the free, massless, integer spin $j=s+1$ (~$h_{\a(s+1)\ad(s+1)}\sim \left[\D_{(\a_{s+1}},\Dd_{(\ad_{s+1}}\right]H_{\a(s))\ad(s))}|$~). From equation (\ref{FC}) we get
\Ibea{l}
\N^{\textit{min}~(1,1)(S,S)}_{\a(s+1)\ad(s+1)} \sim (-i)^{s}\sum_{p=0}^{s}(-1)^{p}\binom{s}{p}^2
\left\{\vphantom{\frac12}
~i~\pa^{(p)}\phi~\pa^{(s+1-p)}\bar{\phi}~-~i\left[\tfrac{2s+1-p}{p+1}\right]~\pa^{(p+1)}\phi~\pa^{(s-p)}\bar{\phi}~\right.\n\label{bosoniccurrent}\\
\hspace{39ex}\left.\vphantom{\frac12}
+\left[ \tfrac{s+p+2}{p+1} \right]~\pa^{(p)}\chi~\pa^{(s-p)}\bar{\chi}~-~\left[\tfrac{s-p}{p+1}\right]~\pa^{(p+1)}\chi~\pa^{(s-p-1)}\bar{\chi}~\right\}~~.
\Ieea
Observe, that there are two contributions into these integer spin currents.
The first one is the boson - boson contribution
and includes the two terms of the first line, where $\phi=\Phi|$.
This corresponds to the bosonic integer spin current
that appears in \cite{current2} and also the traceless part of the currents
in \cite{current1,current6}.
The second contribution is the fermion - fermion one and includes the two terms of the second line
where $\chi_{\a}=\D_{\a}\Phi|$. This corresponds to the fermionic integer
spin current that appears in \cite{current2}.

Furthermore, equation (\ref{f}) gives the conservation of the half-integer spin current $\N^{\textit{min}~(1,0)(S)}_{\a(s+1)\ad(s)}$.
The cubic interactions we get are:
\Ibea{l}
\int d^4x \sum_{s=0}^{\infty} \psi^{\a(s+1)\ad(s)}\N^{\textit{min}~(1,0)(S)}_{\a(s+1)\ad(s)}+c.c.\n
\Ieea
where $\psi_{\a(s+1)\ad(s)}$ is the symmetric, traceless and $\gamma$-traceless part of the free, massless, half-integer
spin $j=s+1/2$ (~$\psi_{\a(s+1)\ad(s)}\sim\left\{\D_{(\a_{s+1}},\Dd^2\right\}H_{\a(s))\ad(s)}|$~). Again using (\ref{FC}) we get
\Ibea{l}
\N^{\textit{min}~(1,0)(S)}_{\a(s+1)\ad(s)}\sim(-i)^{s}\sum_{p=0}^{s}~(-1)^{p}\binom{s}{p}^2\left(\frac{s+1}{p+1}\right)~\pa^{(p)}\chi~\pa^{(s-p)}\bar{\phi}\n\label{fermioniccurrent}~~.
\Ieea
This is the half-integer spin current and appears for the first time in the literature and it has only one contribution of the fermion - boson type.

Finally, we notice that equation (\ref{00}) is the conservation of another current. This corresponds to the $\mathcal{R}$-symmetry
current and it has the form
\Ibea{l}
\N^{\textit{min}~(0,0)}_{\a(s)\ad(s)}\sim(-i)^{s}\sum_{p=0}^{s}~(-1)^{p}\binom{s}{p}^2\left\{~\pa^{(p)}\phi~\pa^{(s-p)}\bar{\phi}+
i\left(\frac{s-p}{p+1}\right)~\pa^{(p)}\chi~\pa^{(s-p-1)}\bar{\chi}~\right\}\n\label{Rcurrent}~~.
\Ieea
%%%%%%%%%%%%%%%%%%%%%%%%%%%%%%%%%%%%%%%%
%%%%%%% Massive Chiral %%%%%%%%%%%%%%%%%
\section{Massive chiral superfield}
\subsection{Higher spin supercurrent and supertrace}
So far we have discussed the higher spin supercurrent multiplet of a free, massless chiral superfield. In this section,
we repeat the analysis for a massive chiral superfield, with a starting action $S_{o}+S_{m}$ where $S_o$ is given by (\ref{S0}) and
$S_{m}$ is the mass term:
\Ibea{l}
S_{m}=\tfrac{m}{2}\int d^6z~\Phi^2~+c.c.\n
\Ieea
The variation of this extra term under (\ref{tr2}) is
\Ibea{l}\n
\d_gS_m=-gm\sum_{k=0}^{\infty}\int d^6z\left\{\vphantom{\frac{1}{2}}\Dd^2\ell^{\a(k+1)\ad(k)}~i^{k}\pa^{(k)}\D\Phi~\Phi~+c.c.\right.\\
\left.\hspace{27ex}-~\Dd^{(\ad_{k+1}}\ell^{\a(k+1)\ad(k))}~i^{k+1}\pa^{(k+1)}\Phi~\Phi~+c.c.
\vphantom{\frac{1}{2}}\right\}\\
\hspace{8ex}+gm\int d^6z~\Dd^2\ell~\Phi~\Phi~+c.c.~\hspace{50ex}.
\Ieea
It is straight forward to show that:
\Ibea{l}
\Dd^2\ell^{\a(k+1)\ad(k)}~i^{k}\pa^{(k)}\D\Phi~\Phi~-~\Dd^{(\ad_{k+1}}\ell^{\a(k+1)\ad(k))}~i^{k+1}\pa^{(k+1)}\Phi~\Phi=\n\vspace{2ex}\\
\hspace{15ex}=~\tfrac{1}{2}\Dd^2\left[\vphantom{\frac{1}{2}}
\Dd^2\ell^{\a(k+1)\ad(k)}~\left\{i^{k}\pa^{(k)}\D\Lambda~\Phi~+~i^{k}\pa^{(k)}\D\Phi~\Lambda\right\}
\right.\\
\hspace{24ex}\left.\vphantom{\frac{1}{2}}
-\Dd^{(\ad_{k+1}}\ell^{\a(k+1)\ad(k))}\left\{i^{k+1}\pa^{(k+1)}\Lambda~\Phi~+~i^{k+1}\pa^{(k+1)}\Phi~\Lambda\right\}\right]~,\\
\Dd^2\ell\Phi~\Phi=\Dd^2\left[\Dd^2\ell~\Lambda~\Phi\right]\n
\Ieea
and by absorbing the overall $\Dd^2$ factor, we can convert the integration over the entire superspace:
\Ibea{l}\n\label{dgSm}
\d_gS_m=\tfrac{g}{2}m\int\sum_{k=0}^{\infty}\left\{\vphantom{\frac{1}{2}}\Dd^2\ell^{\a(k+1)\ad(k)}\left[i^{k}\pa^{(k)}\D\Lambda~\Phi
~+~i^{k}\pa^{(k)}\D\Phi~\Lambda\right]~+c.c.\right.\\
\left.\hspace{21ex}-~\Dd^{(\ad_{k+1}}\ell^{\a(k+1)\ad(k))}\left[i^{k+1}\pa^{(k+1)}\Lambda~\Phi~+~i^{k+1}\pa^{(k+1)}\Phi~\Lambda\right]~+c.c.
\vphantom{\frac{1}{2}}\right\}\\
\hspace{8ex}-gm\int~\Dd^2\ell~\Lambda~\Phi~+c.c.~\hspace{50ex}.
\Ieea
From this expression we can extract the contribution of the mass term to equations (\ref{M},~\ref{N})
However, in order to couple the theory purely to higher spin supermultiplets
the coefficient of $\Dd^{(\ad_{k+1}}\ell^{\a(k+1)\ad(k))}$ must be written as a real term plus total spinorial or spacetime derivative terms. For the massless theory, we have proven this property via equation (\ref{pureHS}) and it holds for any value of $k$. The story for a massive chiral is different as we will show that only the even values of $k=2l$ can satisfy such a requirement.

The relevant quantity for the mass term is $i^{k+1}\pa^{(k+1)}\Lambda~\Phi+i^{k+1}\pa^{(k+1)}\Phi~\Lambda$. It is easy to show that this combination can be written in the following manner:
\Ibea{l}
i^{k+1}\pa^{(k+1)}\Lambda~\Phi~+~i^{k+1}\pa^{(k+1)}\Phi~\Lambda=\vspace{2ex}\\
=
\begin{cases}
i\pa~\left[~\sum\limits_{n=0}^{2l}~(-1)^{l+n}~\pa^{(n)}\Lambda~\pa^{(2l-n)}\Phi~\right]~,~\textit{for}~k=2l~,~l=0,1,2,\dots\vspace{2ex}\n\\
\hphantom{i}\pa~\left[~\sum\limits_{n=0}^{l}~(-1)^{l+n+1}~\pa^{(n)}\Lambda~\pa^{(2l+1-n)}\Phi~+\sum\limits_{n=l+1}^{2l+1}(-1)^{l+n}~\pa^{(n)}\Lambda~\pa^{(2l+1-n)}\Phi~\right]+2~\pa^{(l+1)}\Lambda~\pa^{(l+1)}\Phi~,
\end{cases}
\\ {\vphantom{c}}\hspace{70ex}\textit{for}~k=2l+1~,~l=0,1,2,\dots
\Ieea
therefore, for odd values of $k$ and due to the presence of the term $\pa^{(l+1)}\Lambda~\pa^{(l+1)}\Phi$, there is no 
improvement term $W_{\a(2l+2)\ad(2l+1)}$ to eliminate the imaginary part of $\N_{\a(2l+2)\ad(2l+2)}$. Hence, in order
to construct an invariant theory of a massive chiral interacting with irreducible higher spin supermultiplets, all terms 
in $\delta_{g}\left(S_o+S_m\right)$ that correspond to
an odd value of $k$ must be set to zero. For that reason the parameters $\ell$ and $\ell_{\a(2l+2)\ad(2l+1)}$ for $l=0,1,2,\dots$ 
must vanish and the transformation of $\Phi$ we must consider in this massive case is reduced to:
\Ibea{ll}\n\label{massivetr}
\d_{g}\Phi=&-g\sum_{l=0}^{\infty}\left\{\vphantom{\frac{1}{2}}~\Dd^2\ell^{\a(2l+1)\ad(2l)}~\D_{\a_{2l+1}}\Dd_{\ad_2l}\D_{\a_2l}\dots\Dd_{\ad_1}\D_{\a_1}\Phi\right.\\
&\hspace{9ex}\left.\vphantom{\frac{1}{2}}-\tfrac{1}{(2l+1)!}\Dd^{(\ad_{2l+1}}\ell^{\a(2l+1)\ad(2l))}~\Dd_{\ad_{2l+1}}\D_{\a_{2l+1}}\dots\Dd_{\ad_1}\D_{\a_1}\Phi\right\}
\Ieea

Moreover, we can show that for the case of $k=2l$ the quantity $i^{2l}\pa^{(2l)}\D\Lambda~\Phi
+i^{2l}\pa^{(2l)}\D\Phi~\Lambda$ which appears in (\ref{dgSm}) as the coefficient of $\Dd^2\ell^{\a(2l+1)\ad(2l)}$
can be expressed in the following way:
\Ibea{l}
i^{2l}\pa^{(2l)}\D\Lambda~\Phi+i^{2l}\pa^{(2l)}\D\Phi~\Lambda=
\D\left[~(-1)^{l}~\Lambda~\pa^{(2l)}\Phi~\right]+\pa~\left[~\sum_{n=0}^{2l-1}~(-1)^{l+n+1}~\pa^{(n)}\D\Lambda~\pa^{(2l-1-n)}\Phi~\right]\n
\Ieea
With all the above into account, we get that
\Ibea{ll}\n
\N_{\a(2l+1)\ad(2l+1)}&=X^{(2l+1)}_{\a(2l+1)\ad(2l+1)}+\tfrac{1}{(2l+1)!^2}\Dd_{(\ad_{2l+1}}\D_{(\a_{2l+1}}Z^{(2l+1)}_{\a(2l))\ad(2l))}
-\tfrac{im}{2(2l+1)!^2}\pa_{(\a_{2l+1}(\ad_{2l+1}}Y_{\a(2l))\ad(2l))}~~~~~~\\
&~+\tfrac{1}{(2l+1)!}\D_{(\a_{2l+1}}\Dd^2\bar{U}_{\a(2l))\ad(2l+1)}+\tfrac{2l+1}{(2l+2)(2l+1)!}\Dd_{(\ad_{2l+1}}W_{\a(2l+1)\ad(2l))}~~~.
\Ieea
with
\Ibea{l}
Y_{\a(2l)\ad(2l)}=\sum\limits_{n=0}^{2l}~(-1)^{l+n}~\pa^{(n)}\Lambda~\pa^{(2l-n)}\Phi\n
\Ieea
Now it is obvious that we can always make $\N_{\a(2l+1)\ad(2l+1)}$ real by selecting $W_{\a(2l+1)\ad(2l)}$ as follows:
\Ibea{ll}
W_{\a(2l+1)\ad(2l)}=&-\tfrac{2l+2}{2l+1}\D^2U_{\a(2l+1)\ad(2l)}-\tfrac{2l+2}{(2l+1)(2l+1)!}\D_{(\a_{2l+1}}\hspace{-0.5ex}
\left[Z^{(2l+1)}_{\a(2l))\ad(2l)}
-\tfrac{m}{2}\left(Y_{\a(2l))\ad(2l)}+\bar{Y}_{\a(2l))\ad(2l)}\right)\right]~~~~~~~\n
\Ieea
and the expressions for $\N_{\a(2l+1)\ad(2l+1)}$ and $\M_{\a(2l+1)\ad(2l)}$ become
\Ibea{l}\n
\N_{\a(2l+1)\ad(2l+1)}=X^{(2l+1)}_{\a(2l+1)\ad(2l+1)}+\tfrac{m}{2(2l+1)!^2}\left[\vphantom{\frac12}
\Dd_{(\ad_{2l+1}}\D_{(\a_{2l+1}}\bar{Y}_{\a(2l))\ad(2l))}
-\D_{(\a_{2l+1}}\Dd_{(\ad_{2l+1}}Y_{\a(2l))\ad(2l))}\right]~~~~\n\label{msupcur}
\\
\hspace{15ex}~+\tfrac{1}{(2l+1)!}\left[\vphantom{\frac12}
\D_{(\a_{2l+1}}\Dd^2\bar{U}_{\a(2l))\ad(2l+1)}-\Dd_{(\ad_{2l+1}}\D^2U_{\a(2l+1)\ad(2l))}\right]~~,\vspace{2ex}\\
%%%%%%
\M_{\a(2l+1)\ad(2l)}=\tfrac{1}{(2l+1)!}\D_{(\a_{2l+1}}\M_{\a(2l))\ad(2l)}~~~,\n\vspace{2ex}\\
%%%%%%
\M_{\a(2l)\ad(2l)}=(-1)^{l}\pa^{(2l)}\Phi~\bar{\Phi}-\tfrac{2(l+1)}{2l+1}Z^{(2l+1)}_{\a(2l)\ad(2l)}+\tfrac{m(l+1)}{2l+1}\left(Y_{\a(2l)\ad(2l)}
+\bar{Y}_{\a(2l)\ad(2l)}\right)+\tfrac{m}{2}\Omega_{\a(2l)\ad(2l)}\n\label{msuptr}\\
\hspace{12ex}+\tfrac{2(l+1)}{2l+1}\D^{\a_{2l+1}}U_{\a(2l+1)\ad(2l)}+\Dd^{\ad_{2l+1}}\bar{U}_{\a(2l)\ad(2l+1)}~~
\Ieea
where
\Ibea{l}
\Omega_{\a(2l)\ad(2l)}=(-1)^{l+1}~\Lambda~\pa^{(2l)}\Phi+i\sum_{n=0}^{2l-1}~(-1)^{l+1+n}~\pa^{(n)}\Dd\D\Lambda~\pa^{(2l-1-n)}\Phi~~~.\n
\Ieea

The result for the variation of the $S_o +S_m$ theory is
\Ibea{ll}\n
\d_g\left(S_o+S_m\right)=&-g\int\sum_{l=0}^{\infty}\left\{\vphantom{\frac{1}{2}}\left[\Dd^2\ell^{\a(2l+1)\ad(2l)}-\D_{\a_{2l+2}}\lambda^{\a(2l+2)\ad(2l)}\right]\D_{\a_{2l+1}}\mathcal{T}_{\a(2l)\ad(2l)}+c.c.\right.\\
&\hspace{12ex}\left.\vphantom{\frac{1}{2}}-\tfrac{1}{(2l+1)!}\Dd^{(\ad_{2l+1}}\ell^{\a(2l+1)\ad(2l))}~\mathcal{J}_{\a(2l+1)\ad(2l+1)}+c.c.\right\}
\Ieea
where $\N_{\a(2l+1)\ad(2l+1)}$ and $\M_{\a(2l+1)\ad(2l)}$ are given by (\ref{msupcur}, \ref{msuptr}).
Therefore to get the invariant theory
we have to add  the following higher spin, cubic interaction terms
\Ibea{l}\n
S_{\text{HS-massive chiral}}=g\int\sum_{l=0}^{\infty}\left\{\vphantom{\frac{1}{2}}~~
H^{\a(2l+1)\ad(2l+1)}~\N_{\a(2l+1)\ad(2l+1)}\right.\n\label{mhsci}\\
\hspace{29ex}+\left.\chi^{\a(2l+1)\ad(2l)}~\D_{\a_{2l+1}}\M_{\a(2l)\ad(2l)}+c.c.
\vphantom{\frac{1}{2}}\right\}
\Ieea
Apart from the various mass terms that deform the expressions for the higher spin supercurrent and supertrace, the biggest difference
from the massless chiral story is that the massive chiral superfields has cubic interactions only with $(2l+2~,~ 2l+3/2)$ supermultiplets
that correspond to superspin $Y=2l+3/2$. This includes supergravity $(l=0)$ but not the vector supermultiplet.
\subsection{Minimal multiplet of higher spin supercurrents}
Similar to the massless case, expressions (\ref{msupcur}, \ref{msuptr}) include an arbitrary improvement term
$U_{\a(2l+1)\ad(2l)}$, hence we have to check whether this freedom can be used to completely eliminate the supertrace. For the case of supergravity the \emph{canonical} supercurrent multiplet we get is:
\Ibea{l}
\N_{\a\ad}=\D_{\a}\Phi~\Dd_{\ad}\bar{\Phi}+\tfrac{m}{2}\Dd_{\ad}\D_{\a}\left(\bar{\Lambda}\bar{\Phi}\right)
-\tfrac{m}{2}\D_{\a}\Dd_{\ad}\left(\Lambda\Phi\right)+\D_{\a}\Dd^2\bar{U}_{\ad}-\Dd_{\ad}\D^2 U_{\a}~,\n\\
\M=-\Phi\bar{\Phi}+\tfrac{m}{2}\Lambda\Phi+m\bar{\Lambda}\bar{\Phi}+2\D^{\a}U_{\a}+\Dd^{\ad}\bar{U}_{\ad}~~.\n
\Ieea
It is easy to see that there is no choice of $U_{\a}$ that can cancel the terms of $\M$ proportional to the mass. This is true not just
for the case of supergravity, but for the higher spin supermultiplets as well. The higher spin supertrace $\M_{\a(2l)\ad(2l)}$ can not be eliminated
and there is no \emph{minimal} supercurrent multiplet for massive chirals.

However, we can use the procedure of \textsection \ref{Redef} in order to absorb all the $m$ independent terms of the supertrace and
make it proportional to the mass. In this configuration the supercurrent will be the same as the minimal supercurrent of massless
chiral (\ref{minimalcurrent}) plus terms proportional to mass. For the case of supergravity this will give
\Ibea{l}
\N_{\a\ad}=\N^{\textit{min}}_{\a\ad}-\tfrac{m}{6}\D_{\a}\Dd_{\ad}\left(\Lambda\Phi\right)+\tfrac{m}{6}\Dd_{\ad}\D_{\a}\left(\bar{\Lambda}\bar{\Phi}\right)\n~~,\\
\M=\tfrac{m}{6}\Lambda\Phi+\tfrac{m}{3}\bar{\Lambda}\bar{\Phi}~~.\n
\Ieea
where $\N^{\textit{min}}_{\a\ad}$ is given in (\ref{minimalSG}).
\subsection{Conservation equation}
The conservation equation that the $\N_{\a(2l+1)\ad(2l+1)}$ and $\M_{\a(2l)\ad(2l)}$ satisfy on-shell is
\Ibea{l}
\Dd^{\ad_{2l+1}}\N_{\a(2l+1)\ad(2l+1)}=\tfrac{1}{(2l+1)!}\Dd^2\D_{(\a_{2l+1}}\M_{\a(2l))\ad(2l)}~~,~l=0,1,2,\dots\n\label{mce}
\Ieea
and it is straight forward to show that expressions (\ref{msupcur}, \ref{msuptr}) do that on-shell
\footnote{Keep in mind that the on-shell equation of motion for a free massive chiral is $\Dd^2\bar{\Phi}=m\Phi$~.}. As we did for the massless chiral, we will use this conservation equation to derive a closed form expression for the higher spin supercurrent and supertrace.
Based on the previous results the general ansatz for the higher spin supercurrent and supertrace is
\Ibea{l}
\N_{\a(s)\ad(s)}=\N^{\textit{min}}_{\a(s)\ad(s)}+m\sum_{p=0}^{s-1}\gamma_{p}~\pa^{(p)}\D\Dd\Lambda~\pa^{(s-1-p)}\Phi
+m\sum_{p=0}^{s-1}\delta_{p}~\pa^{(p)}\Dd\Lambda~\pa^{(s-1-p)}\D\Phi\n\label{fmsc}\\
\hspace{21ex}-m\sum_{p=0}^{s-1}\gamma^{*}_{p}~\pa^{(p)}\Dd\D\bar{\Lambda}~\pa^{(s-1-p)}\bar{\Phi}
-m\sum_{p=0}^{s-1}\delta^{*}_{p}~\pa^{(p)}\D\bar{\Lambda}~\pa^{(s-1-p)}\Dd\bar{\Phi}\vspace{1ex}\\
%%%%
\M_{\a(s-1)\ad(s-1)}=~m\sum_{p=0}^{s-1}\zeta_{p}~\pa^{(p)}\Lambda~\pa^{(s-1-p)}\Phi
+m\sum_{p=0}^{s-1}\xi_{p}~\pa^{(p)}\bar{\Lambda}~\pa^{(s-1-p)}\bar{\Phi}\n\label{fmst}\\
\hspace{15ex}+m\sum_{p=0}^{s-2}\sigma_{p}~\pa^{(p)}\Dd\D\Lambda~\pa^{(s-2-p)}\Phi
\Ieea
with $\N^{\textit{min}}_{\a(s)\ad(s)}$ given by (\ref{FC}). The conservation equation (\ref{mce})
fixes the coefficients $\delta_p,~\xi_{p},~\zeta_{p},~\sigma_{p}$:
\Ibea{l}\n
\delta_{p}=-\gamma_{p}~~~~~~~~~~~~~~~~~~~~~~~~~~~~~~~~~p=0,1,\dots,s-1~~,\sn\\
\xi_{p}=-\tfrac{s+1}{s}~\gamma^{*}_{p}~~~~~~~~~~~~~~~~~~~~~~~~~~~p=0,1,\dots,s-1~~,\sn\label{xi}\\
\zeta_{0}=-\tfrac{1}{s}~\gamma_{0}~,\sn\\
\zeta_{p}=-\tfrac{p+1}{s}~\gamma_{p}+\tfrac{s-p}{s}~\gamma_{p-1}~~~~~~~~~~~~p=1,2,\dots,s-1~~,\sn\\
\sigma_{0}=-\tfrac{i}{s}~\gamma_{1}+i~\tfrac{s-1}{s}~\gamma_{0}~,~\sn\\
\sigma_{p}=~(-1)^{p+1}~\tfrac{i}{s}~\gamma_{1}+(-1)^{p}~i~\tfrac{s-1}{s}~\gamma_{0}
+i\sum_{n=1}^{p}~(-1)^{p+n+1}\left[\vphantom{\frac12}
\tfrac{n+1}{s}~\gamma_{n+1}-\tfrac{s-2n-1}{s}~\gamma_{n}-\tfrac{s-n}{s}~\gamma_{n-1}
\right]~,~\sn\\
\hspace{73ex}~~p=1,2,\dots,s-2
\Ieea
and the coefficients $\gamma_{p}$ satisfy the constraints:
\Ibea{l}\n
\gamma_{p}+\gamma_{s-p-1}=\frac{(-1)^{s+p}~(i)^{s+1}}{\binom{2s+1}{s+1}}\sum_{n=0}^{p}\binom{s}{n}^2\left[\vphantom{\frac12}
\tfrac{s+1}{s+1-n}+(-1)^{s}~\tfrac{s+1}{n+1}\right]~,~p=0,1,\dots,s-1\sn\label{gc1}\\
\sigma_{s-2}=-i~\tfrac{s-1}{s}~\gamma_{s-1}+\tfrac{i}{s}~\gamma_{s-2}~,~\sn\label{gc2}
\Ieea
Notice that the left hand side of (\ref{gc1}) is invariant under $p\to s-1-p$, therefore we get a consistency condition
\Ibea{l}
\left[\vphantom{\frac12}~ 1+(-1)^s~\right]\sum_{n=0}^{s}\binom{s}{n}^2\frac{s+1}{n+1}=0\n
\Ieea
which selects only the odd values of $s$, in agreement with (\ref{mhsci}). For $s=2l+1$, equation (\ref{gc1}) fixes $\gamma_{l}$
\Ibea{l}
\gamma_{l}=\frac{l+1}{\binom{4l+3}{2l+2}}\sum_{n=0}^{l}\binom{2l+1}{n}^2\left[\vphantom{\frac12}
\tfrac{1}{2l+2-n}-\tfrac{1}{n+1}\right]\n
\Ieea
A consequence of that is $\xi_{l}\neq0$ due to (\ref{xi}). Therefore the supertrace can not be zero as in the massless case.
Moreover, the constraints (\ref{gc1}, \ref{gc2}) provide a system of $l+2$ linear equations for the $2l+1$, $\gamma_{p}$ coefficients, so there is
a freedom of choice for $l-1$ of these coefficients. This freedom corresponds to the fact that there is no unique canonical supercurrent multiplet, in contrast with the massless case where the minimal multiplet is unique. An example of a choice is to have
\Ibea{l}
\gamma_{l+2}=\gamma_{l+3}=\dots=\gamma_{2l}=0 ~.\n
\Ieea
%%%%%%%%%%%%%%%%%%%%%%%%%%%%%%%%%%%%%%%%
%%%%%%%%%%% Discussion %%%%%%%%%%%%%%%%%%%%%%%
\section{Summary and discussion}
Let us briefly summarize and discuss the results obtained. In
\textsection\ref{dPhi} we presented the most general ansatz for the
transformation of a 4D, ${\cal N}=1$ chiral superfield with linear
terms (\ref{tr}). The consistence with chirality, constrained the
parameters (\ref{chiralconst}) and revealed structures similar to
the gauge transformations of free, massless, higher-superspin
theories. This was a hint that chiral superfields can have cubic
interactions with higher spin superfields. Therefore, using
(\ref{tr2}) and Noether's method we:
\begin{enumerate}
\item[\emph{i})] Proved that a single, massless, chiral superfield can have cubic interactions (\ref{ci})
only with the half-integer superspin $(s+1,s+1/2)$ irreducible representations
of the super-Poincar\'{e} group. Moreover, despite the fact that there are two different formulations
of the half-integer superspin supermultiplets, the chiral superfield has a clear preference to couple
only to one of them, the one that can be lifted to $\mathcal{N}=2$ higher spin supermultiplets.
\item[\emph{ii})] Generated the \emph{canonical} multiplet of higher spin supercurrents $\left\{\N_{\a(k+1)\ad(k+1)},\M_{\a(k)\ad(k)}\right\}$
(\ref{N3}, \ref{M3}) which satisfy conservation equation (\ref{ce1}) and leads to the cubic interactions
\Ibea{l}
g\int\sum_{k=0}^{\infty}\left\{\vphantom{\frac{1}{2}}
H^{\a(k+1)\ad(k+1)}\mathcal{J}_{\a(k+1)\ad(k+1)}
+\chi^{\a(k+1)\ad(k)}\D_{\a_{k+1}}\mathcal{T}_{\a(k)\ad(k)}+\bar{\chi}^{\a(k)\ad(k+1)}\Dd_{\ad_{k+1}}\bar{\mathcal{T}}_{\a(k)\ad(k)}\vphantom{\frac{1}{2}}\right\}~~~~\n\label{cubint2}
\Ieea
The objects $\N_{\a(k+1)\ad(k+1)}$ and $\M_{\a(k)\ad(k)}$ are the higher spin supercurrent and higher spin supertrace respectively and are the higher spin analogues of the supercurrent and supertrace that appear in supergravity.
\item[\emph{iii})] Proved that for every $k$, there is a unique alternative multiplet of higher spin supercurrents, called \emph{minimal}
$\left\{\N^{\textit{min}}_{\a(k+1)\ad(k+1)}~,~0\right\}$ (\ref{minimalcurrent}, \ref{FC}) with conservation equation (\ref{mince}). 
The cubic interactions for the minimal multiplet have the simpler form
\Ibea{l}
g\int\sum_{k=0}^{\infty}~
H^{\a(k+1)\ad(k+1)}\mathcal{J}^{\textit{min}}_{\a(k+1)\ad(k+1)}~~~~.\n\label{minicubint2}
\Ieea
Furthermore,
we presented the construction of the appropriate improvement term that will take us from the \emph{canonical} to the \emph{minimal} multiplet.
The supercurrent $\N^{\textit{min}}_{\a(k+1)\ad(k+1)}$ matches exactly the supercurrent generated by superconformal higher spins presented in \cite{KMT}. 
\end{enumerate}
The identification of the \emph{minimal} multiplet with the results in \cite{KMT} was expected because superconformal higher spin description does not include a compensator like $\chi_{\a(k+1)\ad(k)}$,
hence the cubic interaction terms of the chiral with the superconformal higher spin supermultiplets can only take the form of (\ref{minicubint2}).
However, the superfield $H_{\a(k+1)\ad(k+1)}$ that appears in \cite{KMT} is not the same because its dynamics involve higher derivative terms and also has different engineering dimensions.

In section 9, we discuss the component structure of the theory and specifically we searched for the higher spin currents
generated by the supercurrents. Starting from the superspace conservation equation we project down to the component level
and we find:
\begin{enumerate}
\item[\emph{iv})] An expression for the integer spin current $\mathcal{J}^{\textit{min}~(1,1)(S,S)}_{\a(s+1)\ad(s+1)}$~(\ref{bosoniccurrent}). There are two contributions to this current.
The first is of the boson - boson type constructed out of a complex scalar $\phi$ which is defined as the the $\theta$ independent term of $\Phi$ ($\phi=\Phi|$). The second contribution is of the fermion - fermion type and is constructed out of a spinor $\chi_{\a}$ defined as the $\theta$ term of $\Phi$ ($\chi_{\a}=\D_{\a}\Phi$). Both of these contributions agree with known results.
\item[\emph{v})] An expression for the half-integer spin current $\mathcal{J}^{\textit{min}~(1,0)(S)}_{\a(s+1)\ad(s)}$ (\ref{fermioniccurrent}). This current appears for the first time in the literature because it requires both the complex scalar and the spinor, therefore
non-supersymmetric theories can not be used to construct it.  
\item[\emph{vi})] An expression for an $\mathcal{R}$-symmetry current $\mathcal{J}^{\textit{min}~(0,0)}_{\a(s)\ad(s)}$ (\ref{Rcurrent}). This current also appears for the first time.
\end{enumerate}

It is important to emphasize that in general the higher spin supercurrent and higher spin supertrace are independent quantities
and the $\emph{minimal}$ multiplet can not always be reached. It depends on the peculiarities of the starting action and 
its symmetries, such as superconformal, to decide whether this can be done or not. In this work, we present a method of constructing the higher spin supercurrent and supertrace which is not restricted by these considerations.
In section 10, we discuss the higher spin supercurrent multiplet
of a massive chiral superfield. Our results are:
\begin{enumerate}
\item[\emph{vii})] A massive chiral can have cubic interactions only with the odd $s$ $[s=2l+1]$ half-integer superspin supermultiplets $(2l+2~,~2l+3/2)$. 
\item[\emph{viii})] The expressions for the higher spin supercurrent $\N_{\a(2l+1)\ad(2l+1)}$ (\ref{msupcur}, \ref{fmsc}) and supertrace $\M_{\a(2l)\ad(2l)}$ (\ref{msuptr}, \ref{fmst}) of the \emph{canonical} multiplet. These expressions have not been obtained before.
\item[\emph{ix})] There is no \emph{minimal} multiplet of supercurrents for this case since the supertrace can not be adsorbed by improvement terms. However, it can be arranged to be proportional to the mass parameter, so at the massless limit we land at the \emph{minimal} multiplet of the massless chiral superfield.
\end{enumerate}

There are several directions for the further development and
generalization of the superfield interaction vertices studied in the
paper. Firstly, the approach under consideration can directly be
applied to derivation of the cubic interaction of the
higher-superspin superfield with chiral superfield on the $AdS$
superspace background. Secondly, it would be extremely interesting to
construct the supercurrent and corresponding cubic interaction
vertex for 4D, ${\cal N}=2$ massless higher-superspin gauge
superfield. In this case the supercurrent should apparently be built
from hypermultiplet superfields on the framework of harmonic
superspace \cite{GIOS} which provides unconstrained superfiled
description for 4D, ${\cal N}=2$ supermultiplets. Thirdly, it would be
interesting to apply this approach to other matter supermultiplets 
such as the complex linear.

{\bf Acknowledgements}\\[.1in] \indent
The authors are thankful to G. Tartaglino-Mazzucchelli for participation 
in the early stages of this work. Also the authors want to thank S.\ M.\ Kuzenko
for extremely useful discussions and M. Taronna for correspondence. I.\ L.\ B\
is grateful to the RFBR grant, project No. 15-02-03594-a and to Russian
Ministry of Education and Science, project No. 3.1386.2017 for partial support.
The work of K.\ K.\ was supported by the grant P201/12/G028 of the Grant
agency of Czech Republic.

%%%%%%%%%%%%%%%%%%%%%%%%%%%%%%%%%%%%
%%%%%%%%% Bibliography %%%%%%%%%%%%%%%%%%%%

\end{document}